\documentclass[
    aps,
    prb,
    reprint,
    superscriptaddress,
    twocolumn,
    amsmath,
    amssymb,
    longbibliography,
    floatfix
]{revtex4-2}

\usepackage{
    amsfonts,
    bbm,
    bbold,
    braket,
    comment,
    dsfont,
    graphicx,
    mathrsfs,
    multirow,
    physics
}
\usepackage[normalem]{ulem}
\usepackage[mathscr]{euscript}
\usepackage{xcolor}
\usepackage[
    colorlinks,
    linkcolor=blue,
    citecolor=blue,
    urlcolor=blue
]{hyperref}
\usepackage{orcidlink}
\usepackage{microtype}
\usepackage{bm}
\usepackage[capitalise]{cleveref}

\newcommand{\bk}{\mathbf{k}}
\newcommand{\pf}{\operatorname{Pf}}
\newcommand{\id}{\mathbbm{1}}

\newcommand{\A}{\mathcal{A}}
\newcommand{\Z}{\mathbb{Z}}

\newcommand{\sgn}{\mathrm{sgn}}

\newcommand{\K}{\mathcal{K}}

\begin{document}
\title{Majorana flat bands from mirror--particle--hole antisymmetry}

\author{Justin Schirmann\orcidlink{0009-0007-7030-0155}}
\email{jschirm@sandia.gov}
\affiliation{Center for Integrated Nanotechnologies, Sandia National Laboratories, Albuquerque, New Mexico 87185, USA}

\author{Martin Rodriguez-Vega\orcidlink{0000-0001-8929-6546}}
\affiliation{American Physical Society, 1 Physics Ellipse Dr, College Park, MD, USA}

\author{Alexander Cerjan\orcidlink{0000-0002-4362-7300}}
\affiliation{Center for Integrated Nanotechnologies, Sandia National Laboratories, Albuquerque, New Mexico 87185, USA}
\date{\today}

\begin{abstract}
Majorana flat bands, zero modes across the whole edge Brillouin zone, usually rest on chiral symmetry.
We show that a fully gapped class-D superconductor can host such a flat band without chiral symmetry or
time reversal, provided its superconducting gap is odd under a mirror and translation along the edge is
preserved. Combined with particle--hole conjugation, the mirror forms an antiunitary antisymmetry that
squares to $+1$ and preserves the momentum along the edge, so it fixes the Kitaev index of every slice of
fixed edge momentum. In a gapped bulk, this slice index cannot change with the edge momentum, and the
edge zero modes form a band
across the whole edge Brillouin zone. We construct a four-band model with exactly these symmetries, a
helical $p$-wave superconductor with Rashba coupling in an in-plane exchange field, from which no term can
be removed without restoring an additional symmetry. On the mirror lines, its two mirror sectors are exact Kitaev chains. This yields a closed-form
phase diagram. Besides the flat band, the model hosts a phase with two Majoranas at each end of a wire,
although its class-D index is trivial, and a second-order phase with Majorana corner modes. The spectral localizer detects all these indices in real space, directly on finite-size samples as relevant for experiments. On open rectangular flakes of two widths, it
reproduces the phase diagram even though translational symmetry is broken. In the nodal regions, its reading depends on the flake
size, so they are left unread. The readings persist under disorder several times the bulk gap.
Without translation along the edge, the symmetries protect no
more than Majorana parities, and we quantify the signatures that remain specific to the antisymmetry. A
minimal model of magnetic adatoms on a triplet superconductor realizes the flat-band and second-order
phases.
\end{abstract}

\maketitle

\section{Introduction}
\label{sec:intro}

Topological superconductors bind Majorana zero modes to their boundaries and
defects~\cite{ReadGreen2000, Kitaev2001}. A pair of these self-conjugate modes stores one
fermion nonlocally, and their non-Abelian exchange statistics make them candidate building blocks for
fault-tolerant quantum computation~\cite{Kitaev2003, Nayak2008}. The search for platforms in which such
modes can be created and braided has moved from
proposals at topological-insulator interfaces and in semiconductor wires~\cite{FuKane2008,
Sau2010, Lutchyn2010, Oreg2010} to magnetic chains and islands on
superconductors~\cite{NadjPerge2014, Menard2017, Kezilebieke2020} and other engineered
platforms~\cite{Alicea2012, SatoAndo2017, FlensbergvonOppenStern2021}. Energetically, particle--hole symmetry,
present in the BdG description of a superconductor, pins an isolated Majorana mode at zero energy; spatially, which boundaries and
defects carry such modes is set by the system's dimension and by the other symmetries. The tenfold way assigns invariants to each class of time-reversal,
particle--hole and chiral symmetry~\cite{AltlandZirnbauer1997, SchnyderRyuFurusakiLudwig2008,
Kitaev2009, ChiuTeoSchnyderRyu2016}, and superconductors with particle--hole symmetry alone form
class D. A fully gapped two-dimensional class-D superconductor is characterized by
the Chern number $C\in\Z$ of its Bogoliubov--de Gennes Hamiltonian and hosts $|C|$ chiral Majorana modes
at its edge~\cite{ReadGreen2000}. Specific system geometries can also bind Majoranas at particle--hole-invariant edge
momenta, such as weak indices found in coupled wires~\cite{SeroussiBergOreg2014}. Crystalline symmetries
refine this classification~\cite{ChiuYaoRyu2013, MorimotoFurusaki2013, ShiozakiSato2014}: they
can protect several Majoranas per edge~\cite{TewariSau2012, ZhangKaneMele2013, FangGilbertBernevig2014,
PhongSainzCruzMeleGuinea2024}, Majoranas bound to the corners of higher-order
phases~\cite{BenalcazarBernevigHughes2017, Schindler2018, SongFangFang2017, Langbehn2017,
Geier2018}, and Majorana \emph{flat bands}, zero modes across the whole edge Brillouin zone.

Flat bands are the extreme case of boundary zero modes. Their extensive degeneracy gives a zero-bias conductance that
grows with the edge length and makes the edge susceptible to interactions~\cite{TanakaKashiwaya1995,
PotterLee2014}. They were first predicted as zero-energy Andreev states at the surfaces of $d$-wave
superconductors~\cite{Hu1994, TanakaKashiwaya1995}, where time reversal combined with particle--hole
symmetry is a chiral symmetry. More generally, Majorana flat bands rest on a chiral symmetry, whether it
arises from time reversal or from another antiunitary symmetry~\cite{RyuHatsugai2002,
SatoTanakaYadaYokoyama2011, TanakaSatoNagaosa2012}. At each momentum
along the edge, chiral symmetry defines a winding number of the Bloch Hamiltonian along the momentum
normal to the edge, and a nonzero winding binds zero modes at that edge momentum. The winding can change
only where the bulk gap closes, so in nodal superconductors the flat bands span the regions between
projected bulk nodes~\cite{SchnyderRyu2011, SchnyderBrydon2015, MatsuuraChangSchnyderRyu2013, BrydonSchnyderTimm2011,
SchnyderBrydonTimm2012, WongLiuLawLee2013, HyartWrightRosenow2014, ChenLiuXuChen2023,
NagaeKatayamaIkegaya2025}, including nodes protected by mirrors~\cite{ChiuSchnyder2014,
KobayashiSumitaYanaseSato2018}. In a gapped bulk, the winding is the same at every edge momentum. If it is nonzero, the flat band
covers the whole edge Brillouin zone. Examples are weakly coupled chains~\cite{SenguptaZuticKwonYakovenkoDasSarma2001},
gapped bulks with an extra spin-rotation symmetry~\cite{LappLinkTimm2024}, and lattices whose
inhomogeneous magnetic fields leave an effective chiral operator~\cite{SedlmayrAguiarHualdeBena2015}. Without chiral symmetry,
a reflection combined with particle--hole symmetry can pin Majorana arcs between projected point
nodes~\cite{ChiuSchnyder2014}, flat bands can appear along vortex lines~\cite{ZhangDoubleMajorana2026,
ZhangMa2026}, and mirrors of Shiba lattices can protect edge and corner modes~\cite{Soldini2023}.

What has been missing is a mechanism that pins a Majorana flat band across the whole edge Brillouin
zone of a fully gapped class-D superconductor, without chiral symmetry or time reversal. Without a
mirror, the weak index of a class-D superconductor binds Majoranas only at the edge momenta $0$ and
$\pi$~\cite{SeroussiBergOreg2014}. With a mirror, Chiu and Schnyder showed that the reflection combined
with particle--hole symmetry acts as an effective particle--hole symmetry of each slice of fixed edge
momentum, which protects Majorana arcs between projected nodes in nodal
superconductors~\cite{ChiuSchnyder2014}. If this symmetry protection could be extended
across the entire edge Brillouin zone of a fully gapped superconductor, it would establish a new mechanism
for realizing extensive Majorana zero modes without chiral symmetry or time reversal.

In this work, we show that a mirror combined with particle--hole conjugation can produce a Majorana
flat band across the whole edge Brillouin zone. In nodal superconductors, this combination acts as an
effective particle--hole symmetry within each slice of fixed edge momentum and protects Majorana
arcs~\cite{ChiuSchnyder2014}. Once the bulk is fully gapped, the arcs become a flat band, provided the
superconducting gap is odd under the mirror and translation along the edge is preserved. The protecting operation, the mirror combined with
particle--hole conjugation, yields an antiunitary antisymmetry $\A$ that squares to $+1$ and fixes the Kitaev
index of every slice, so that the weak index of class D becomes a flat band. A gap odd under a mirror is thus a straightforwardly checkable criterion for a
Majorana flat band in a fully gapped superconductor without chiral symmetry. We construct a minimal four-band model in this class, a helical
$p$-wave superconductor with Rashba coupling in an in-plane exchange field, whose two mirror sectors are
exact Kitaev chains on the mirror lines. This gives a closed-form phase diagram that contains, besides
the flat band, a phase with two Majoranas at each end of a wire although its class-D index is trivial,
and an intrinsic second-order phase with Majorana corner modes. We then read these phases in real space with
the spectral localizer~\cite{Loring2015, LoringSchulzBaldes2017}, which computes invariants directly
from the Hamiltonian and position operators of a finite sample and therefore applies to the small flakes
in which such lattices are assembled. It recovers the whole phase diagram from open flakes alone and follows each
index into the disordered regime. Finally, we determine what protects each phase beyond the clean limit
and present a magnetic-adatom model that realizes the flat-band and second-order phases.

The paper is organized as follows. \Cref{sec:symmetry} constructs $\A$ and shows that it turns the weak
index into a flat band. To show that the mechanism is realized, \cref{sec:model} builds a minimal model
with exactly these symmetries. Because the index of $\A$ detects only the flat band, \cref{sec:sectors}
resolves the mirror sectors of the model and obtains its full phase diagram. Since the lattices of interest are finite flakes, \cref{sec:realspace} reads all these indices in real
space. \Cref{sec:protection} asks which indices and which boundary modes survive disorder. \Cref{sec:adatoms}
proposes a magnetic-adatom realization, and \cref{sec:conclusion} concludes. The appendices contain
derivations and numerical checks.

\section{Symmetry and flat band}
\label{sec:symmetry}

We first ask what it takes to pin a Majorana at every edge momentum of a gapped two-dimensional class-D
superconductor. Particle--hole symmetry alone pins them only at the edge momenta $0$ and $\pi$: any other
slice of fixed $k_y$ is mapped onto the slice at $-k_y$ and has no index of its own. Thus, pinning a Majorana at
every $k_y$ requires an antiunitary operation that acts within each slice and squares to $+1$. In this section, we construct such an operation from a mirror and a particle--hole conjugation and
identify the crystalline class it defines. We then show that, in a fully gapped bulk, it turns the weak
index of class D into a flat band. We also derive zero-dimensional indices on the lines $k_x=0,\pi$, which force nodes
where they change. The index protected by $\A$ is a single
sign per slice, as it detects the flat band but no other phase, which is a limitation that \cref{sec:sectors}
lifts.

With $\sigma_i,\tau_i$ Pauli matrices in spin and Nambu space and $\K$ complex conjugation, the
particle--hole operation $\Xi=\tau_x\K$, $\Xi^2=+1$, is present in the BdG description of a superconductor and cannot be broken; $\K$ also reverses
$\bk$. We
compose a spinful mirror about $y$ with a particle--hole operation dressed by a $\pi$ shift of
the pairing phase,
\begin{equation}
\label{eq:factors}
    M_y = i\sigma_yP_y,\ \ \Xi' = i\tau_y\K = \tau_z\Xi,\ \ M_y^2=\Xi'^2=-1,
\end{equation}
with $P_y$ the spatial reflection $y\to-y$ (in momentum space, $k_y\to-k_y$). Their product is the protecting operation,
\begin{equation}
\label{eq:A}
    \A = M_y\Xi' \simeq \tau_y\sigma_y\,\K\,P_y,
    \qquad
    \A^2=+1.
\end{equation}
Both $\K$ and $P_y$ reverse $k_y$, so $\A$ reverses $k_x$ alone and acts \emph{within} each $k_y$
slice. That $\A$ squares to $+1$ is essential: for $\A^2=-1$ each slice would be in class C, whose
one-dimensional index is trivial~\cite{SchnyderRyuFurusakiLudwig2008, Kitaev2009}. A spin-only
mirror gives $M_y^2=-1$, and $\tau_z$, which generates a $\pi$ shift of the condensate phase,
supplies the compensating sign: $\Xi'$ is particle--hole conjugation combined with the gauge
rotation $e^{i\pi\tau_z/2}=i\tau_z$, which shifts the pairing phase by $\pi$. The
product
\begin{equation}
\label{eq:W}
    W = \Xi\A = i\tau_z\sigma_yP_y,
    \qquad
    W^2=-1,\ \ \{W,\Xi\}=0,
\end{equation}
is a unitary reflection mapping $k_y\to-k_y$. Since $iW$ is Hermitian and commutes with $\Xi$, the system is in class D with a reflection $\mathcal R_+$.
Its crystalline classification~\cite{ChiuYaoRyu2013, MorimotoFurusaki2013, ShiozakiSato2014} in two
dimensions is a strong $\Z_2$, carried by Majorana corner states~\cite{Langbehn2017, Geier2018,
TrifunovicBrouwer2019}, plus weak indices of stacked chains. This is the entry $\mathcal R_+$, $\Z_2$ of
class D in Table~I of Ref.~\cite{Langbehn2017}, where $\mathcal R_+$ squares to $+1$ and commutes with
particle--hole conjugation, as $iW$ does here. Since $\Xi$ is always present, $\A$ is
present exactly when $W$ is: physically, the requirement is a mirror under which the gap is
odd~\cite{UenoYamakageTanakaSato2013, ZhangKaneMele2013}. The operation $\A$ is the effective
particle--hole symmetry of Chiu and Schnyder~\cite{ChiuSchnyder2014}, whose $\Z_2$ index protects point
nodes away from the mirror lines; combinations of crystalline operations with particle--hole conjugation
are reviewed in Ref.~\cite{MizushimaEtAl2016}.

\Cref{tab:mech} compares this mechanism with earlier ones.

\begin{table}[t]
    \caption{Mechanisms for Majorana flat bands and arcs, condensed from \cref{tab:mechanisms}. Time reversal underlies the chiral symmetry of the first and third rows; this work needs neither.}
    \label{tab:mech}
    \begin{ruledtabular}
    \small
    \renewcommand{\arraystretch}{1.25}
    \begin{tabular}{p{0.95in}p{0.9in}p{0.42in}p{0.7in}}
        System & Symmetry & Bulk & Edge zero modes \\
        \hline
        $d$-wave, nodal~\cite{Hu1994, TanakaKashiwaya1995, SchnyderRyu2011} & chiral & nodal & arcs between nodes \\
        \hline
        Reflection-symmetric nodal~\cite{ChiuSchnyder2014} & reflection with $\Xi$ (or chiral) & nodal & arcs between nodes \\
        \hline
        Coupled chains~\cite{SenguptaZuticKwonYakovenkoDasSarma2001} & chiral & gapped & whole edge zone \\
        \hline
        Weak class D~\cite{SeroussiBergOreg2014} & $\Xi$ only & gapped & $k_y=0,\pi$ only \\
        \hline
        This work & $\A=\Xi W$ (mirror-odd gap) & gapped & whole edge zone \\
    \end{tabular}
    \end{ruledtabular}
\end{table}

Within each $k_y$ slice, $\A$ acts as the particle--hole symmetry of a one-dimensional class-D chain.
Its Kitaev index $\nu(k_y)$ is therefore defined at every $k_y$, not only on the mirror lines, and can
change only where the slice closes its gap. In nodal superconductors it changes across the projected nodes,
and the edge zero modes form arcs between them~\cite{ChiuSchnyder2014}. In a fully gapped bulk it
cannot change, so it equals the weak index, which we denote $\nu\equiv\nu(k_y)$. If $\nu=-1$, every slice binds one Majorana at each
end of a ribbon such that the zero modes form a band pinned at $E=0$ across the edge Brillouin zone. The band
needs translation along the edge, which keeps $k_y$ a good quantum number; without it, only the
parity of the Majorana number at an end survives (\cref{sec:protection}). For $k_y \notin \{0,\pi\}$, $\Xi$ exchanges the zero modes at $k_y$ and $-k_y$. Each of them is
self-conjugate under $\A$, the particle--hole symmetry of its slice. In this sense, the band consists entirely of
Majorana modes, just like the flat bands of $d$-wave superconductors.

The two lines $k_x^*\in\{0,\pi\}$ carry more information. There $\A$ maps every point $(k_x^*,k_y)$ to itself, so $H(k_x^*,k_y)U_\A$ is antisymmetric at every
$k_y$, with $U_\A=\tau_y\sigma_y$ the matrix part of $\A$, which is real and symmetric. The sign of the Pfaffian of $H$, whose square is $\det H$, is a
zero-dimensional class-D index along the whole line~\cite{ChiuSchnyder2014}. The Kitaev index $\nu(k_y)$
is the product of these signs on the two $k_x^*$ lines.  If
this sign differs between $k_y=0$ and $\pi$, the spectrum must close on that line: these are the point
nodes of Ref.~\cite{ChiuSchnyder2014}.

\section{Minimal model}
\label{sec:model}

In \cref{sec:symmetry} we showed that a superconducting gap odd under the mirror $y\to-y$ yields an
antisymmetry $\A$ that, in a fully gapped bulk, pins a Majorana at every edge momentum. Whether a
concrete Hamiltonian realizes such a flat band, and which other phases can accompany it, requires a model,
which also makes every claim explicitly checkable. We look for the simplest Hamiltonian with exactly the symmetries of \cref{sec:symmetry} and no
others, so that no additional symmetry can be responsible for the flat band. We work on a square lattice with one site per unit cell. Among the sixteen on-site and nearest-neighbor terms that these symmetries allow, we select eight that
leave only $\{\id,W,\Xi,\A\}$ intact. Removing any one of them restores an additional symmetry. The
result is a helical $p$-wave superconductor with Rashba coupling in an in-plane exchange field. This model's Kitaev index $\nu$ has a closed form,
its flat band is flat to machine precision and persists when all sixteen terms are present, and it is
adiabatically connected to an array of Kitaev chains.

\begin{figure}[t]
    \centering
    \includegraphics[width=\columnwidth]{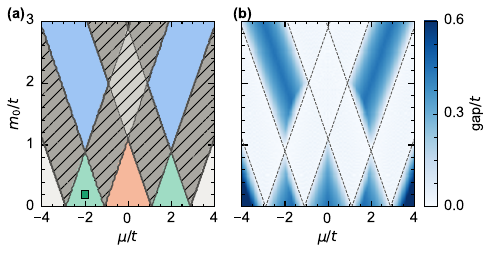}
    \caption{Phases of the minimal model Hamiltonian, \cref{eq:model}, at hopping amplitudes $t=1$ (along $x$) and $t_y=0.45$ (along $y$), $p_x$ pairing $a=0.8$, Rashba coupling $b_1=0.3$ and $p_y$ pairing $b_2=1$, in the plane of chemical potential $\mu$ and in-plane exchange field $m_0$. Region names are defined in \cref{sec:sectors} and \cref{tab:phases}. (a) The
    sector indices given by \cref{eq:nusector} denote a flat band ($\nu=-1$, blue), $(-,-)$ (orange), double flip (green), trivial (white), sector
    swap (light gray) and single flip (dark gray). Hatching denotes a bulk gap below $0.03$, which marks the two
    nodal regions. Square indicates $(\mu,m_0)=(-2,0.2)$, used in
    \cref{fig:corners}. (b)~Bulk gap; dashed lines indicate $|\mu\mp m_0|=2t\pm2t_y$.}
    \label{fig:phases}
\end{figure}

\begin{figure*}[t]
    \centering
    \includegraphics[width=\textwidth]{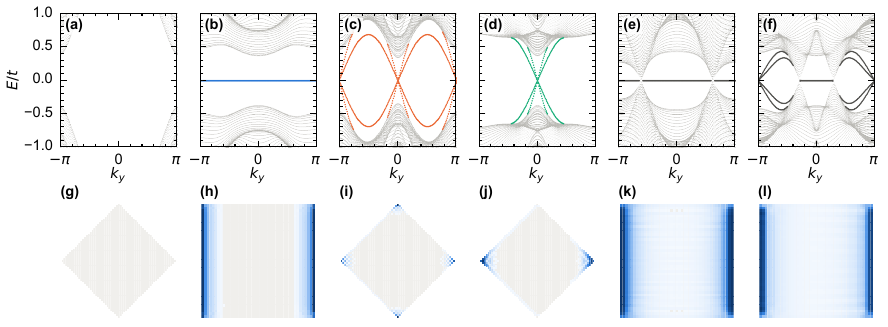}
    \caption{Boundary states of every phase of \cref{fig:phases}(a), at $(\mu,m_0)$ and with bulk gap:
    trivial (a,g), $(3.5,0.2)$, $0.39$; flat band (b,h), $(2.4,2)$, $0.36$; $(-,-)$ (c,i), $(0.3,0.2)$, $0.34$;
    double flip (d,j), $(-2,0.2)$, $0.36$; sector swap (e,k), $(0.3,2.2)$, nodal; and single flip
    (f,l), $(1.2,0.8)$, nodal; $t=1$, $t_y=0.45$, $a=0.8$, $b_1=0.3$, $b_2=1$. (a)--(f)~Ribbon open along $x$ ($N_x=60$); colored: more than half the weight within six
    sites of the edges. In (b) the residual splitting of the flat band, from hybridization of the two
    edges across the ribbon, is $6\times10^{-11}$; it falls to $4\times10^{-15}$ for $N_x=120$. (g)--(l)~Zero-energy local density of states
    $\sum_n|\psi_n(\bm r)|^2\eta/\pi(E_n^2+\eta^2)$, $\eta=0.005$, on the diamond $|x|+|y|\le30$, except in
    (h), (k) and (l), a $43\times43$ square whose $x$-normal edges carry the flat band and the Majorana arcs (its $y$-normal edges are gapped); normalized per panel. The numbers of states with $|E|<0.02$ in (g)--(l) are $0$, $86$, $8$, $2$, $84$
    and $26$; on a flake they are not
    protected individually (\cref{sec:protection}).}
    \label{fig:densities}
\end{figure*}

A search with Qsymm~\cite{Varjas2018} over point-group, time-reversal, particle--hole and chiral
operations, together with a direct search over momentum shifts (\cref{app:model}), confirms that
$\{\id,W,\Xi,\A\}$ is the full symmetry group of the minimal model,
\begin{equation}
\begin{split}
    \label{eq:model}
    H(\bk) =& \epsilon_\bk\tau_z + m_0\sigma_y
    + a\sin k_x\,\tau_x\sigma_x
    \\&+ b_1\big(\sin k_y\,\tau_0\sigma_x - \sin k_x\,\tau_z\sigma_y\big)
    + b_2\sin k_y\,\tau_y\sigma_0,
\end{split}
\end{equation}
with
\begin{equation}
\epsilon_\bk=-2t\cos k_x-2t_y\cos k_y-\mu.
\end{equation}
Here $t$ and $t_y$ are the nearest-neighbor hoppings along $x$ and $y$, $\mu$ is the chemical potential, $m_0$ is an in-plane exchange field along $y$, $a$ is the equal-spin $p_x$ pairing, $b_1$ is the Rashba coupling, and $b_2$ is the $p_y$ triplet pairing. Hopping, an in-plane exchange field, and
equal-spin $p_x$ pairing form a spinful Kitaev array that still has the chiral symmetry
$\tau_x\sigma_z$; a Rashba
coupling removes it, and a $p_y$ triplet removes a residual $C_{2z}\mathcal T$. The model is minimal in its class, because removing any one of the eight terms restores an additional
symmetry. Removing $m_0$ restores time reversal, and removing the $\sin k_x$ Rashba term restores the
chiral operation. Removing $a$, the $\sin k_y$ Rashba term or $b_2$ restores a $C_{2z}\mathcal T$-type
antiunitary. Removing $\mu$, $t$ or $t_y$ restores a sublattice operation combined with a momentum
shift (\cref{app:model}). Only
the triplets break $M_y$ and $\Xi'$ individually; $\A$ and $\Xi$ are exact. We fix the couplings $(t,t_y,a,b_1,b_2)$, whose values are given in each
figure caption, and vary $(\mu,m_0)$. The other parameter sets used for checks
are listed in \cref{tab:params}.

\Cref{fig:phases}(a) contains six regions, named in \cref{sec:sectors} and listed in \cref{tab:phases}. Here we only need $\nu$, the index protected by $\A$. For \cref{eq:model} the index $\nu$ has a closed form (\cref{app:invariants}), so the phase diagram that $\A$ alone
predicts is analytic. The index is $\nu=-1$ in the flat-band and sector-swap regions of \cref{fig:phases}(a). In the sector-swap region, the mirror sector that is inverted at $k_y=0$ is the opposite one from that inverted at $k_y=\pi$. The bulk is then nodal, because the zero-dimensional index of \cref{sec:symmetry} changes sign between $k_y=0$ and $\pi$ along $k_x=0$ (\cref{sec:nodal}), so the edge carries Majorana arcs rather than a flat band.
Where $\nu$ differs between $k_y=0$ and $\pi$, the bulk must close its gap. At all other gapped points
$\nu=+1$, and $\A$ alone predicts no edge zero modes. Where $\nu=-1$, every $k_y$ slice is an inverted Kitaev chain, and a ribbon open along $x$ carries a
zero-energy band [\cref{fig:densities}(b)]. Any residual splitting seen in a finite system comes from the hybridization of the
two edges across the ribbon and decays exponentially with $N_x$. The flatness holds at every gapped point with $\nu=-1$ and is lost only when $\A$ is broken. For
example, $\lambda\tau_y\sigma_y$, which preserves $\Xi$, gaps the band linearly as
$0.37\lambda|\sin k_y|$ (\cref{app:invariants}). Zeros then remain only at $k_y=0,\pi$, where the weak indices of
$\Xi$ apply.

The flat band does not rely on the residual $C_{2z}\mathcal T$ of the
model at $b_2=0$. Instead, it stays flat when $C_{2z}\mathcal T$ is broken and $\A$ kept, and gaps linearly when
an in-plane Zeeman field along $x$ breaks $\A$ but keeps $C_{2z}\mathcal T$ (\cref{app:invariants}). The term
$b_2$ is needed only to remove the residual $C_{2z}\mathcal T$ and for the gap of the double flip below.

The results do not rely on the minimal form. We add the eight other allowed terms with random
coefficients up to $0.15$. The bulk gap at $(2.4,2)$ is then $0.49$, and the ribbon band is flat to
$2\times10^{-12}$. The indices still follow from the terms even in $k_x$, which are the only ones that
survive at $k_x^*$ (\cref{app:invariants}). At the four particle--hole-invariant momenta, $\pf(HU_\A)=\pf(H\tau_x)$ holds
identically for \cref{eq:model} and in every random model of this class that we tested (\cref{app:invariants}). For these models, the remaining weak index of class D is $+1$ whenever the bulk is gapped, and the
Chern number is even. This is the weak index of chains along $y$,
$\nu_y(k_x^*)=\sgn[\pf(H(k_x^*,0)\tau_x)\,\pf(H(k_x^*,\pi)\tau_x)]$.

The flat-band phase is adiabatically connected to a familiar limit. Along the path
$(b_1,b_2)=u\,(0.3,1)$, $0\le u\le1$, at $(\mu,m_0)=(2.4,2)$ the bulk gap decreases monotonically from
$0.56$ to $0.36$ without closing. At $u=0$ the model is a spin-polarized array of $p_x$ Kitaev chains
with the chiral operation $\tau_x\sigma_z$, and its flat band is the chiral one of weakly coupled
chains~\cite{SenguptaZuticKwonYakovenkoDasSarma2001, SatoTanakaYadaYokoyama2011}. The Rashba coupling
breaks the chiral operation; $\A$ is what keeps the band flat once it does.

\section{Mirror sectors and phase diagram}
\label{sec:sectors}

In the minimal model of \cref{sec:model}, \cref{eq:model}, $\nu=-1$ only in the flat band parameter regime among the gapped regions; every other gapped region has
$\nu=+1$ and $\A$ alone says nothing about it. Nevertheless, the mirror resolves these regions, as on the mirror lines $k_y\in\{0,\pi\}$, the Hamiltonian splits into two mirror sectors,
each with its own particle--hole symmetry. In this section, we show that for the model of \cref{sec:model}, each sector is
exactly a Kitaev chain, such that four sector Majorana numbers, and with them the whole phase diagram,
follow in closed form. The result is four gapped phases---a flat band, a first-order phase with two
Majoranas per end, a second-order phase with Majorana corner modes, and a trivial phase---and two nodal
regions. We compute the boundary states of each and explain them with a simple picture of stacked
Kitaev chains.

As $\Xi$ is antiunitary and anticommutes with $W$, it maps each $W$ eigenspace to itself, so each
mirror sector carries its own particle--hole symmetry. At transverse momentum $k_y\in\{0,\pi\}$, in
the eigenbasis of $-iW=s$ (Pauli matrices $\rho_i$), each sector of \cref{eq:model} is exactly a
Kitaev chain~\cite{Kitaev2001},
\begin{equation}
\label{eq:sectorH}
    h_s(k_x; k_y) = \xi_s(k_x; k_y)\,\rho_z + a\sin k_x\,\rho_y - s\,b_1\sin k_x\,\rho_0,
\end{equation}
with
\begin{equation}
\xi_s (k_x; k_y)=-2t\cos k_x-(\mu-sm_0+2t_y\cos k_y).
\end{equation}
In \cref{eq:sectorH}, the exchange field splits the chemical potentials
of the two chains and the Rashba term tilts them oppositely. The tilt vanishes at $k_x\in\{0,\pi\}$, so the
sector Majorana numbers $\nu_s^{k_y}$ follow from
\begin{equation}
\label{eq:nusector}
    \nu_s^{k_y} = -1 \iff |\mu-sm_0+2t_y\cos k_y| < 2t,\quad k_y\in\{0,\pi\},
\end{equation}
with $\nu(k_y)=\nu_+^{k_y}\nu_-^{k_y}$, so that $\nu(0)=\nu(\pi)=\nu$ in every gapped phase. Sector $s$ is therefore inverted at both momenta if
$|\mu-sm_0|<2t-2t_y$, at neither if $|\mu-sm_0|>2t+2t_y$, and at $0$ ($\pi$) only if
$|\mu-sm_0\pm2t|<2t_y$, so every phase boundary is one of the lines $|\mu\mp m_0|=2t\pm2t_y$
[\cref{fig:phases}]. Like any class-D index, $\nu_s$ is stable under adding trivial bands, so none
of these phases are fragile (\cref{app:model}). The four sector indices $\nu_\pm^{0,\pi}$ define four gapped phases and two nodal regions. The gapped
phases are the flat band (a weak phase), $(-,-)$ (first-order crystalline), the double flip
(second-order) and the trivial phase. The nodal regions are the sector swap and the single flip. Altogether, the pattern of
inverted $W$ sectors that defines each possible phase is summarized in \cref{tab:phases}.
The names refer to which sectors are inverted on which mirror line,
relative to the mirror about a site row; \cref{fig:densities} shows the
boundary states of each.

\begin{table}[t]
    \caption{Sign patterns of the $W$-sector Majorana numbers of \cref{eq:nusector} that define the six
    regions of \cref{fig:phases}(a), relative to the site-row mirror; $\nu=\nu_+^k\nu_-^k$ at $k=0,\pi$ and
    $\chi=\nu_+^0\nu_+^\pi$ [\cref{eq:chi}]. Each pattern is given up to exchanging $s=\pm$, and for the
    double and single flip also up to exchanging $k_y=0$ and $\pi$. The sector swap and the single flip
    are nodal.}
    \label{tab:phases}
    \begin{ruledtabular}
    \begin{tabular}{lcccl}
        Region & $(\nu_+^0,\nu_-^0,\nu_+^\pi,\nu_-^\pi)$ & $\nu$ & $\chi$ & Majoranas \\
        \hline
        Trivial & $(+,+,+,+)$ & $+1$ & $+1$ & none \\
        Flat band & $(-,+,-,+)$ & $-1$ & $+1$ & flat edge band \\
        $(-,-)$ & $(-,-,-,-)$ & $+1$ & $+1$ & two per end \\
        Double flip & $(-,-,+,+)$ & $+1$ & $-1$ & one per corner \\
        Sector swap & $(-,+,+,-)$ & $-1$ & $-1$ & arcs \\
        Single flip & $(-,+,+,+)$ & $\pm1$ & $-1$ & arcs \\
    \end{tabular}
    \end{ruledtabular}
\end{table}

Each of the $W$-sector-specified phases has a simple real-space picture as stacks of Kitaev chains along $x$ on the
mirror-symmetric rows, in the spirit of block-state constructions of crystalline
phases~\cite{SongHuangFuHermele2017}. A site row is a row of lattice sites along $x$, at integer $y$. A bond row is the line along $x$ halfway between two neighboring site rows, at half-integer $y$. The stack is periodic along $y$, with one site row and one bond row in each unit cell. With $n_{\rm site}$ and $n_{\rm bond}$ the numbers of chains per unit cell on site and on bond rows, counted modulo two,
\begin{equation}
\label{eq:chi}
    \nu=(-1)^{n_{\rm site}+n_{\rm bond}},\qquad \chi\equiv\nu_+^0\nu_+^\pi=(-1)^{n_{\rm bond}},
\end{equation}
with $\chi$ taken relative to a site-row mirror. Physically, $\chi$ is the strong $\Z_2$ index of
class D with $\mathcal R_+$ (\cref{sec:symmetry}). When $\nu=+1$, $\chi=-1$ means that a single Majorana is bound
at each point where the mirror axis meets a gapped edge, as in the double flip below.
Here a chain is a Kitaev chain along $x$ in its topological phase, lying in a single $W$ sector $s$.
\Cref{eq:chi} follows from the contribution of one such chain per cell to the sector indices. Without
interchain coupling, the Hamiltonian of the stack does not depend on $k_y$, so every slice contains one
copy of the chain and has $\nu(k_y)=-1$. Kitaev indices of decoupled blocks multiply, since the Pfaffian
of a direct sum is the product of the Pfaffians. A chain on a site row is mapped to itself by $W$ and lies
in sector $s$ on both mirror lines. It flips $\nu_s^0$ and $\nu_s^\pi$ and leaves $\chi$ unchanged. The
site-row mirror maps a bond row onto the neighboring bond row, so on such a chain $W$ includes a
translation by one lattice spacing, which multiplies its eigenvalue at $k_y=\pi$ by $-1$. The chain then
lies in sector $s$ at $k_y=0$ and in $-s$ at $\pi$, and flips $\chi$. Multiplying the contributions of all
chains gives \cref{eq:chi}. Couplings between the chains cannot change the sector indices without closing
the gap on a mirror line. Moreover, the four stacks with a single chain, on a site or a bond row and in
either sector, generate all eight sign patterns with $\nu(0)=\nu(\pi)$. Every point with a gapped bulk
therefore has the sector indices of some stack, and \cref{eq:chi} holds there. It fails only in the single
flip, where $\nu(0)\ne\nu(\pi)$, and, being a statement modulo two, it does not separate $(-,-)$ from the
trivial phase. The flat band is a stack on site rows, the sector swap the same stack on bond rows, the
double flip has chains on both, and $(-,-)$ has one chain per site row in each sector.

We now discuss the six phases in turn, each defined by which sectors are inverted on which mirror
line. They are the flat band (\cref{sec:flat}), the $(-,-)$ phase with two Majoranas per end
(\cref{sec:mm}), the second-order double flip (\cref{sec:df}) and the two nodal regions
(\cref{sec:nodal}). The trivial phase, in which no sector is inverted, serves as the reference.

\subsection{Flat band}
\label{sec:flat}

In the flat-band phase [\cref{fig:densities}(b,h)], \emph{one sector is inverted on both mirror lines
and the other on neither}. It is the only gapped phase detected by $\nu$, the index that $\A$ protects. Unlike the flat bands of nodal and coupled-chain
superconductors~\cite{SchnyderRyu2011, SenguptaZuticKwonYakovenkoDasSarma2001} or of
time-reversal-invariant ones~\cite{LappLinkTimm2024}, it needs no chiral symmetry, and its bulk is
fully gapped; \cref{tab:mech} compares the protecting symmetries of these mechanisms. The pattern of inverted $W$ sectors that defines this phase,
$(\nu_+^0,\nu_-^0,\nu_+^\pi,\nu_-^\pi)=(-,+,-,+)$ up to exchanging $s=\pm$, is defined relative to $W$,
the mirror about a site row. The mirror about a bond row differs from $W$ by one lattice translation.
That translation multiplies the $W$ eigenvalues at $k_y=\pi$ by $-1$ and so exchanges the two $W$
sectors there. It therefore maps the flat-band pattern onto the sector-swap pattern, $(-,+,+,-)$, and
back. Whether a gapped phase with $\nu=-1$ is called flat band or sector swap thus depends on the choice
of mirror line. It records whether the stacked Kitaev chains sit on site rows or on bond rows. In
\cref{eq:model} the region called sector swap is nodal.

\subsection{Two Majoranas per end}
\label{sec:mm}

When \emph{both sectors are inverted on both lines}, inside the diamond $|\mu|+m_0<2t-2t_y$, the
system is in the $(-,-)$ phase. It has two Majoranas per end of each wire while $\nu=+1$, as in the
chiral, Kramers and mirror multiplets of Refs.~\cite{TewariSau2012, WongLaw2012,
GaidamauskasPaaskeFlensberg2014, ZhangKaneMele2013}. On a ribbon, the two edge modes at each end hybridize away from $k_y \in \{0,\pi\}$, and $W$ pins only their
crossings [\cref{fig:densities}(c)]. Each pinned crossing binds a Majorana where the mirror axis meets a gapped edge~\cite{Langbehn2017,
Geier2018}. Pairs of Majoranas can be removed, locally if they share a sector and by decorating the edges if
not~\cite{TrifunovicBrouwer2019, Khalaf2021}. Only the parity $\tfrac12[1-\nu_+^0\nu_+^\pi]$ of the
number of crossings is therefore a bulk property. With two crossings, the corner modes of $(-,-)$ depend on the corner angle. We test this on wedges $|x|+|y|/\beta\le R$ at $(\mu,m_0)=(0.3,0.2)$ and $R=40$. For $\beta=1$, there are
four corner levels with energies between $3\times10^{-7}$ and $5\times10^{-7}$ (in units of $t$) [as in \cref{fig:densities}(i)]. They do not decrease with
$R$ and grow linearly under a $\Xi$- and $W$-symmetric corner perturbation. For $\beta=2$, each corner
instead binds one zero mode per sector, immune to that perturbation [\cref{fig:corners}(b,c)]. The same
holds at the couplings $t_y=0.15$, $b_2=0.13$ of \cref{tab:params}. The $(-,-)$ phase is thus a first-order crystalline phase, with extrinsic, angle-dependent corner modes (\cref{app:corners}).

\subsection{Second-order phase}
\label{sec:df}

\emph{Both sectors inverted at $k_y=0$ and neither at $\pi$}, or the reverse, gives the
double-flip diamonds $\big||\mu|-2t\big|+m_0<2t_y$. Here $\nu=+1$ at both momenta and $\A$ alone predicts nothing. Nevertheless, as
$\nu_+^0\nu_+^\pi=-1$, the phase exhibits an odd number of pinned crossings, which is the strong $\Z_2$ of class D
with $\mathcal R_+$, and $b_2$ opens
a bulk gap of up to $0.46$ (\cref{app:invariants}). Exactly one Majorana sits at each mirror-axis corner
[\cref{fig:densities}(j) and \cref{fig:corners}(a)] for every angle tested, split by $10^{-14}$ at
$R=40$ and immune to symmetric corner perturbations [\cref{fig:corners}(c)]. This is an intrinsic
second-order topological superconductor~\cite{WangLinHughes2018, YanSongWang2018, WangLiuLuZhang2018}.
At $m_0=0$ the model regains time reversal $i\sigma_y\K$ and is a helical
superconductor~\cite{QiHughesRaghuZhang2009}. The exchange field turns it into the double-flip phase, as
in-plane fields do in helical superconductors and heterostructures~\cite{Zhu2018, Pahomi2020,
WuHouLiLuoShiZhang2020, IkegayaRuiManskeSchnyder2021, ChenLiuXuChen2023}. This phase borders the flat
band.

\subsection{Nodal phases}
\label{sec:nodal}

In the sector swap, inside $|\mu|+|m_0-2t|<2t_y$, \emph{one sector is inverted only at $k_y=0$ and the
other only at $\pi$}. This leaves $\nu=-1$ at both momenta, so $\A$ predicts a flat band, but the bulk
is nodal. In \cref{eq:model}, $\pf[H(0,k_y)U_\A]=m_0^2-\epsilon(0,k_y)^2+(b_1^2-b_2^2)\sin^2k_y$ is
negative at $k_y=0$ and positive at $k_y=\pi$ throughout the swap region, for every $b_1$ and $b_2$ (the Rashba coupling and the $p_y$ pairing, respectively).
The zero-dimensional index of \cref{sec:symmetry} therefore forces a node on the line $k_x=0$, and
likewise on $k_x=\pi$. This relies on the
sector masses of \cref{eq:model}, not on symmetry alone: the adatom lattice of \cref{sec:adatoms} has a
gapped swap without Rashba coupling (\cref{app:shiba}). If \emph{a single sector} is inverted at one mirror line only and the other sector at both or neither
(single flip), $\nu$ differs between $k_y=0$ and $\pi$, which forces a node. In both regions, $\A$ still pins a Majorana at every $k_y$ whose slice is gapped and inverted. The edge
therefore carries Majorana arcs, zero-energy bands that end at the projected nodes
[\cref{fig:densities}(e,f) and (k,l)], as in the nodal superconductors of Ref.~\cite{ChiuSchnyder2014}. On a
flake, the low-energy states of these regions are edge states whose tails reach into the gapless bulk
[\cref{fig:densities}(k,l)]; their number grows with the edge length, not the area.

\section{Real-space indices}
\label{sec:realspace}

\begin{figure*}[t]
    \centering
    \includegraphics[width=\textwidth]{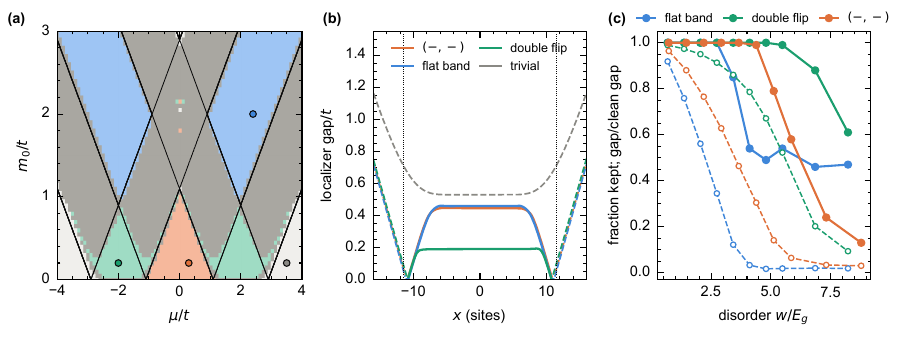}
    \caption{The phase diagram of a finite version of the minimal model given by \cref{eq:model} determined using the spectral localizer, $\kappa=0.15\,t$ per lattice spacing; energies in
    units of $t$; $t_y=0.45$, $a=0.8$, $b_1=0.3$, $b_2=1$. (a)~Phase diagram from open flakes alone: sector localizers at the centre of $24\times10$
    and $24\times11$ flakes. A point is colored flat band if the product of all four signs, $\nu$, is
    $-1$; otherwise double flip if the product on the $24\times10$ flake, $\chi$, is $-1$; otherwise $(-,-)$
    if both signs of the $24\times11$ flake are $-1$, and trivial if not. Colors as in \cref{fig:phases}(a); gray where a
    second pair, $24\times14$ and $24\times15$, gives different signs or a localizer gap (\textit{i.e.} $g_L=\min|\text{spec}(L)|$) is below $0.02$.
    Lines: the Bloch phase boundaries of \cref{fig:phases}(a). Dots: the points of (b) and (c), flat band
    $(2.4,2)$, double flip $(-2,0.2)$, $(-,-)$ $(0.3,0.2)$ and trivial $(3.5,0.2)$. (b)~Gap of the
    particle--hole localizer, \cref{eq:loc}, along $y=0$ of clean flakes ($24\times11$ for the flat band,
    $24\times10$ otherwise), solid where $\zeta(x)=-1$ and dashed where $+1$; for $(-,-)$, the smaller sector-localizer gap on
    the $24\times11$ flake, solid where $\zeta_+=\zeta_-=-1$. Dotted vertical lines: the ends of the flake. (c)~On-site disorder
    $V(\bm r)\tau_z$, $V$ uniform in $[-w,w]$, in units of the bulk gap $E_g$ at each point ($0.36$, $0.36$ and $0.34$): fraction of $100$ samples that
    keep the clean reading (filled symbols, solid lines) and median localizer gap at $x=0$ relative to
    the clean one (open symbols, dashed lines). Flat band:
    $\nu$ from the particle--hole localizer on the two flakes of (a), disorder breaking the mirror; double
    flip: $\zeta_{\rm even}$ on the $24\times10$ flake, likewise; $(-,-)$: sector signs of the $24\times11$ flake,
    mirror-symmetric disorder.}
    \label{fig:localizer}
\end{figure*}

\Cref{sec:symmetry,sec:sectors} characterize these $\mathcal{A}$- and $W$-sector-protected topological phases through the indices $\nu$ and
$\nu_s^{k_y}$, defined in momentum space such that translation symmetry is assumed. However, the lattices that motivate this work, assembled atom by atom~\cite{Kim2018, Schneider2021, Soldini2023}, are small flakes in which the
boundary is a large fraction of the sample and a Brillouin zone is, at best, approximate. We therefore
derive the same indices in real space within the spectral localizer framework. The spectral
localizer~\cite{Loring2015, LoringSchulzBaldes2017, LoringSchulzBaldes2020} computes topological invariants
from the Hamiltonian and position operators in real space~\cite{LoringHastings2010, CerjanLoring2022}. It has been used to compute the invariants of class-D Shiba
lattices~\cite{RodriguezVegaLoringCerjan2025} and weak indices from the parity of the sample
width~\cite{FulgaPikulinLoring2016}. It has also resolved crystalline indices with a unitary
symmetry~\cite{CerjanLoringSchulzBaldes2024} and, through an antiunitary $C_{2z}\mathcal T$, fragile
phases~\cite{Lee2025}. In particular, here we consider a particle--hole spectral localizer that needs no mirror. We derive a
counting rule that relates its sign on flakes of even and odd width to the weak and strong indices. We
then resolve it by mirror sector on flakes and cylinders. Assembled, these readings reproduce the phase diagram of
\cref{fig:phases}(a) from open flakes alone, even though translational symmetry is broken, and recognize the nodal regions by the
dependence of the reading on the flake size [\cref{fig:localizer}(a)].

\begin{figure}[t]
    \centering
    \includegraphics[width=\columnwidth]{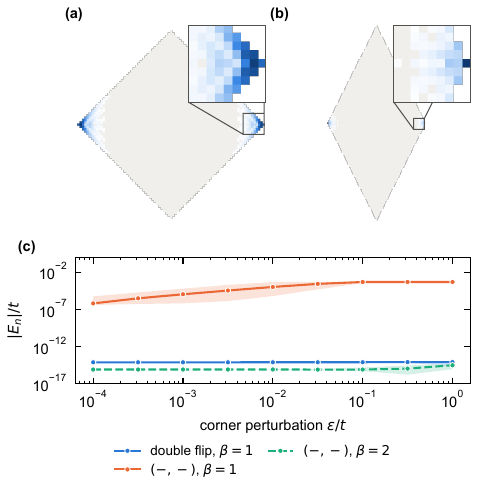}
    \caption{Corner Majoranas on wedges $|x|+|y|/\beta\le R$, $R=40$, at $t=1$, $t_y=0.45$, $a=0.8$,
    $b_1=0.3$, $b_2=1$; square root of the density of the
    lowest states; insets: zoom on the $9\times9$ sites at the corner $(R,0)$. (a)~Two lowest states at
    $(\mu,m_0)=(-2,0.2)$ in the double-flip phase [square in \cref{fig:phases}(a)], $\beta=1$. (b)~Four lowest states at
    $(\mu,m_0)=(0.3,0.2)$ in the $(-,-)$ phase, $\beta=2$. (c)~$|E_n|$ ($n$ the number of corner
    modes) versus the strength of a random $\Xi$- and $W$-symmetric perturbation within three sites of
    $(R,0)$; median and range of five samples, for the double flip at $\beta=1$ (one Majorana per corner)
    and for $(-,-)$ at $\beta=1$ (two in the same sector) and $\beta=2$ (one per sector). In the double flip $|E_2|$ stays constant to machine precision.}
    \label{fig:corners}
\end{figure}

In the Majorana basis where $\Xi$ is complex conjugation, the Hamiltonian $H_M$ of a flake is purely
imaginary and the position $X$ real. Viewing the flake as a wire along $x$, a one-dimensional class-D system, we use the one-dimensional spectral
localizer~\cite{LoringSchulzBaldes2017, DollSchulzBaldes2021},
\begin{equation}
\label{eq:loc}
    L(x) = \kappa(X-x)\otimes\sigma_y+H_M\otimes\sigma_z ,
\end{equation}
which is Hermitian and, since $H_M$ is imaginary and $X$ real, also imaginary and antisymmetric, so its Pfaffian is defined directly. For a flake of $N$
sites, $L$ has dimension $8N$, so $\pf[-iL]=(-i)^{4N}\pf L=\pf L$, and passing to the real skew matrix
$-iL$, which is convenient numerically, adds no sign. The overall sign of the Pfaffian does depend on
the basis, since $\pf(OAO^T)=\det O\,\pf A$, and hence on the choice of Majorana basis and on the
ordering of the tensor products in \cref{eq:loc}. We remove this dependence by normalizing to a
point $x_0$ far outside the flake, where $L(x_0)$ is dominated by the position term and has no
Majorana between it and the flake edge, and define
\begin{equation}
\label{eq:zeta}
    \zeta(x)=\sgn\pf[-iL(x)]\;\sgn\pf[-iL(x_0)] .
\end{equation}
With this definition, $\zeta(x)=-1$ exactly when an odd number of Majoranas, pinned to $E=0$ by
particle--hole symmetry, lies between $x_0$ and $x$. For $x$ inside the flake, $\zeta$ is therefore the
parity of the number of Majoranas at the $x$ end facing $x_0$. The localizer gap $g_L(x)=\min|\mathrm{spec}\,L(x)|$ is the smallest absolute eigenvalue of $L(x)$; $\zeta(x)$ can change sign only where it closes. Unless stated
otherwise, we read $\zeta$ at the center of the flake, $x=0$. In all numerics, $x$ is measured from the center of the flake, which spans $|x|\le(L_x-1)/2$, and $x_0=2L_x+10$. To assess a two-dimensional flake with
the one-dimensional localizer of \cref{eq:loc}, we view it as a wire along $x$ whose $L_y$ rows are
internal degrees of freedom. Each of these rows then contributes a factor $\nu$ to $\zeta$, and each $W$-pinned crossing of the edge
contributes one more Majorana. The first contribution is how a weak index shows up as a dependence on
the parity of the sample width~\cite{FulgaPikulinLoring2016}. The number of crossings is odd exactly when
$\chi=-1$ (\cref{sec:sectors}), so on flakes of even and odd width
\begin{equation}
\label{eq:flakes}
    \zeta_{\rm even}=\chi,\qquad \zeta_{\rm odd}=\nu\,\chi,
    \qquad (\nu,\chi)\in\Z_2\times\Z_2 .
\end{equation}
Two flakes thus give $\nu=\zeta_{\rm odd}\zeta_{\rm even}$ and $\chi$, and a single flake of even width
cannot detect the flat band, as expected for a weak index. The same signs follow from the stacked
Kitaev chains of \cref{sec:sectors} by counting the chain ends at an $x$ edge of the flake. There are
$L_y$ of them in the flat band, $L_y-1$ in the sector swap, $2L_y-1$ in the double flip and $2L_y$ in
$(-,-)$, and their parity reproduces \cref{eq:flakes}. Equivalently, $\zeta=\nu\chi_c$ with $\chi_c$
taken relative to the flake's own mirror line, a site row for odd and a bond row for even $L_y$.
\Cref{eq:flakes} is exact for long flakes. Away from its ends along the $x$-direction, a flake is a strip with
translation along $x$, and $\zeta$ is the Kitaev index of that strip,
$\sgn[\pf(-iH_M(k_x=0))\,\pf(-iH_M(\pi))]$.

Two caveats apply to $\chi$. First, it is a strong index only when $\nu=+1$ (trivial, $(-,-)$ and double
flip): a mirror through bonds instead of sites maps $\chi\to\nu\chi$, so for $\nu=-1$ it records only
whether the chains sit on site or bond rows (\cref{sec:sectors}). Second, reading it requires
mirror-symmetric terminations. A single Kitaev chain attached along one $y$-normal edge adds one
Majorana at each $x$ end and flips $\zeta$ on flakes of both widths without closing the bulk or the
localizer gap. Decorating both edges restores the readings, and $\nu$, read on two identically
terminated flakes, is unaffected (\cref{app:localizer}).

The pair $(\nu,\chi)$ does not separate $(-,-)$ from the trivial phase, since both have $\nu=\chi=+1$.
In $(-,-)$ each $W$ sector carries one Majorana per end, and the two cancel in $\zeta$. Separating the two
phases therefore requires counting the Majoranas of each $W$ sector separately, which the mirror makes
possible. On a flake symmetric under $y\to-y$, conjugation by $\A$ maps $(X,Y,H)\to(+X,-Y,-H)$. This is the
superconducting counterpart of the $C_{2z}\mathcal T$ of Ref.~\cite{Lee2025}, which acts as
$(-X,-Y,+H)$. On such a flake, $W$ commutes with $X$ and $H$, so \cref{eq:loc} restricted to each
sector is again real. Its sign $\zeta_s$, with $s=\pm$, counts the Majoranas of sector $s$ at the end
of the flake, and $\zeta=\zeta_+\zeta_-$~\cite{CerjanLoringSchulzBaldes2024}. A flake still mixes the two mirror lines $k_y=0$ and $\pi$. Resolving them, and with them all four
sector indices of \cref{tab:phases}, requires a cylinder. On a cylinder, open along $x$ and periodic along $y$, $\zeta_s$ separates the two mirror-symmetric
transverse momenta $0$ and $\pi$. For a circumference $L_y\equiv1\pmod4$, $\zeta_s$ is the sector
Majorana number $\nu_s^0$. A second cylinder with $L_y\equiv2\pmod4$ then gives
$\nu_{-s}^\pi=\zeta_s(L_y\equiv1)\,\zeta_s(L_y\equiv2)$. These four real-space indices, derived exactly on cylinders (\cref{app:localizer}), reproduce \cref{eq:nusector}
and separate all six regions. They agree at every bulk-gapped grid point where the localizer gap is
resolved, with the grid statistics given in \cref{app:localizer}. The remaining index of class D, the weak index of chains along $y$, is
trivial in every gapped phase (\cref{sec:model}).

\Cref{fig:localizer} assembles these readings into a phase diagram computed from open flakes alone.
On a $24\times10$ and a $24\times11$ flake, the sector signs give $\nu=\zeta_{\rm odd}\zeta_{\rm even}$
and $\chi=\zeta_{\rm even}$ through \cref{eq:flakes}. When both are $+1$, the sector signs
$(\zeta_+,\zeta_-)=(-,-)$ of the odd flake separate $(-,-)$ from the trivial phase. A finite flake has a
gap even where the bulk has nodes, so the localizer always returns signs; in the nodal regions, however,
they depend on the size of the flake. We therefore keep a reading only where a second pair, $24\times14$
and $24\times15$, of the same widths modulo $4$, returns the same signs and all localizer gaps exceed
$0.02$. The result [\cref{fig:localizer}(a)] does not assume translation invariance. On an $81\times61$ grid, the localizer reads $2003$ of the $2174$ bulk-gapped points and agrees with
\cref{fig:phases}(a) at $1997$ of them. The six misreads lie at $m_0\le0.05$, next to the line $m_0=0$,
where the model is a helical class-DIII superconductor with gapless edges (\cref{app:localizer}). The unread points
lie within two grid steps of a phase boundary or at $m_0\le0.1$. Of the $2767$ points in the nodal regions, $2561$ are left
unread, and $181$ of the other $206$ lie within two grid steps of a gapped phase. Fixing
$L_x=24$ hides no size dependence along $x$: on $32\times10$ and $32\times11$ flakes, the reading is
unchanged at all $205$ nodal points read and at $200$ sampled gapped points. \Cref{fig:localizer}(b)
shows what is read: along the axis of a flake the localizer gap closes where $\zeta(x)$ changes sign,
at the two ends where the Majoranas sit.

The localizer constructions themselves are those of Refs.~\cite{LoringSchulzBaldes2017,
FulgaPikulinLoring2016, CerjanLoringSchulzBaldes2024}; what the model adds is that their readings follow in closed form from the
sector Kitaev chains of \cref{eq:sectorH}, those on flakes through the counting rule of
\cref{eq:flakes}.

\section{Protection beyond the clean limit}
\label{sec:protection}

While \cref{sec:realspace} considered clean flakes, this section asks which of the three nontrivial
phases, and which of their boundary modes, survive disorder. Real superconductors are never clean, and disorder is far from a mere nuisance. By Anderson's
theorem, nonmagnetic disorder leaves the transition temperature of a conventional $s$-wave
superconductor unchanged~\cite{Anderson1959}, and near the superconductor--insulator transition the
multifractality of nearly localized wave functions can even enhance pairing and raise the critical
temperature~\cite{FeigelmanIoffeKravtsovYuzbashyan2007, BurmistrovGornyiMirlin2012}. Unconventional
pairing is far more fragile: in Sr$_2$RuO$_4$, nonmagnetic impurities destroy superconductivity once the
mean free path approaches the coherence length~\cite{Mackenzie1998}. Topology adds a further twist.
Disorder can create topological phases that the clean system lacks, as in the topological Anderson
insulator~\cite{LiChuJainShen2009} and in Rashba wires where scattering induces Majorana end
states~\cite{AdagideliWimmerTeker2014}. Weak disorder can even shorten the Majorana localization length
and make a one-dimensional topological superconductor more robust~\cite{HaimStern2019}. In two
dimensions, class-D disorder produces a thermal metal with delocalized quasiparticles between the
insulating phases~\cite{SenthilFisher2000}. For lattices assembled atom by atom, the question is unavoidable. An adatom array has missing atoms, a
spread of exchange couplings and a substrate with impurities of its own. Which features of a phase
survive this disorder decides what an experiment can see.

 We add on-site disorder $V(\bm r)\tau_z$, with $V$ drawn uniformly from $[-w,w]$, so that $w$
sets the disorder strength. $V$ is drawn either independently on every site, which breaks the mirror,
or symmetrically under $y\to-y$, which respects it. The sector-resolved readings
require disorder that respects the mirror, at least on average over an
ensemble~\cite{FulgaVanHeckEdgeAkhmerov2014}, whereas $\nu$ requires only particle--hole symmetry and is
stable against weak disorder that breaks the mirror (\cref{app:localizer}). We follow two kinds of
quantities: the boundary zero modes, through the spectra and the Andreev conductance of finite flakes,
and the real-space indices of \cref{sec:realspace}, through the localizer readings of
\cref{fig:localizer}(c). The three phases rest on different ingredients and respond differently. The flat band relies on $\A$
and on translation along the edge, the $(-,-)$ phase on $W$, and the corner modes of the double flip on
the gaps of the edges that meet there. For the flat band, we also identify the
signature that is specific to $\A$.

We begin with the flat band of \cref{sec:flat}, whose protection has three parts. The antisymmetry
$\A$ pins the band only while translation along the edge is preserved. Once translation is broken, only
Majorana parities survive, although the weak index read in real space survives disorder of several
times the bulk gap. Finally, the signature specific to $\A$ is a linear splitting of the band under
perturbations that break it.

In a clean ribbon, $\A$ pins the band at zero (\cref{sec:model}). On clean rectangular flakes at $(\mu,m_0)=(2.4,2)$, the $2L_y$ edge levels converge to zero as the
width grows (median $1\times10^{-6}$ at $L_y=21$). The exceptions are eight levels at about $(3$--$5)\times10^{-3}$, localized at the four corners where the flat-band edges meet the gapped
$y$-normal edges. The corners themselves break translation along the edge (\cref{app:transport}). On-site disorder of strength $w=0.2$ splits the edge levels to a
median of $1\times10^{-3}$ whether or not it respects the mirror, and so $\A$. Only Majorana parities survive~\cite{QueirozSchnyder2014, IkegayaKobayashiAsano2018}. The parity
protected by $\Xi$ leaves one Majorana per end for odd $L_y$. With mirror-symmetric disorder, the two
mirror sectors each protect a parity as well, but for the flat band these amount to the same Majorana. The weak index, which assigns one such parity to each row, survives much
stronger disorder, as weak indices generally do as long as the bulk gap stays
open~\cite{RingelKrausStern2012, FulgaPikulinLoring2016}. We read it with the particle--hole localizer on independently disordered $24\times11$ and
$24\times10$ flakes, with disorder that breaks the mirror. Up to $w=1$, about $2.7$ times the bulk gap,
$\zeta_{\rm odd}=-1$ and $\zeta_{\rm even}=+1$ hold in all $100$ samples, so
$\nu=\zeta_{\rm odd}\zeta_{\rm even}=-1$ in every pair [\cref{fig:localizer}(c)]. At $w=1.25$ they hold
in $90$ and $91$ samples, and both are lost only once the localizer gap has closed. The extensive zero-bias
Andreev conductance, $2L_ye^2/h$ on a ribbon, is reduced only gradually. On a $40\times9$ flake contacted along one $x$-normal edge, it falls
from $0.94$ to about $0.85$ of $2L_ye^2/h$ at $w=0.7$, about twice the bulk gap, with or without the
mirror (\cref{app:transport}).
This numerical observation signals the presence of a flat band, but does not by itself test $\A$.

The signature specific to $\A$ is the linear splitting of the band by perturbations that break $\A$, such as a
tilt of the exchange field out of plane by an angle $\theta$, which replaces $m_0\sigma_y$ in
\cref{eq:model} by $m_0(\cos\theta\,\sigma_y+\sin\theta\,\tau_z\sigma_z)$. In \cref{eq:model} a tilt of $5^\circ$ splits the band
by about $0.02\Delta_{\max}$, with $\Delta_{\max}=\max_\bk|\Delta(\bk)|$, about $25$ times the median
disorder splitting at $w=0.2$ (\cref{app:transport}). Interactions that respect $\A$ and translation can still split the clean band by breaking one of them
spontaneously. Edge ferromagnetism splits the flat bands of $d$-wave superconductors in this
way~\cite{PotterLee2014}, and a surface instability with spontaneous spin polarization and currents
splits those of nodal noncentrosymmetric ones~\cite{TimmRexBrydon2015}. We have not studied such effects
here.

The double-flip phase of \cref{sec:df} responds quite differently to disorder, because its corner
Majoranas are single zero modes rather than a band. The mirror places them at the corners
on the mirror axis, but a single Majorana cannot gap locally: it leaves the corner only when the gap of
an edge that meets there closes. Such corner-localized Majoranas are therefore immune to $\Xi$- and $W$-symmetric corner
perturbations [\cref{fig:corners}(c)] and to mirror-symmetric disorder up to $w=1$, and do not split
even when $W$ is broken uniformly up to $\lambda=0.1$ (\cref{app:corners}). Accordingly, the Majorana parity at an end of the even flake, $\zeta_{\rm even}=-1$, survives
mirror-breaking disorder in all $100$ samples up to $w=1.75$, about $4.8$ times the bulk gap (in $99$ at
$w=2$) [\cref{fig:localizer}(c)]. It equals $\chi$ while the mirror is present; once the mirror is broken
it is protected by $\Xi$ alone, as a single Majorana, and no longer by the strong index.

Finally, in the $(-,-)$ phase of \cref{sec:mm}, the two Majoranas at each end of a wire are protected
by $W$, which places one in each sector. With $\Xi$ alone they could pair up, since their total class-D index
is trivial. Chemical-potential disorder commutes with $W$ and keeps all four modes at zero up to about
$2.5$ times the wire gap ($0.45$ for the wire of \cref{fig:protection}). Zeeman disorder that anticommutes
with $W$, by contrast, splits them at any strength (\cref{app:transport}). On flakes, the sector signs $(-,-)$ survive mirror-symmetric
disorder in all $100$ samples up to $w=1.25$, about $3.7$ times the bulk gap [\cref{fig:localizer}(c)].

\begin{figure*}[t]
    \centering
    \includegraphics[width=\textwidth]{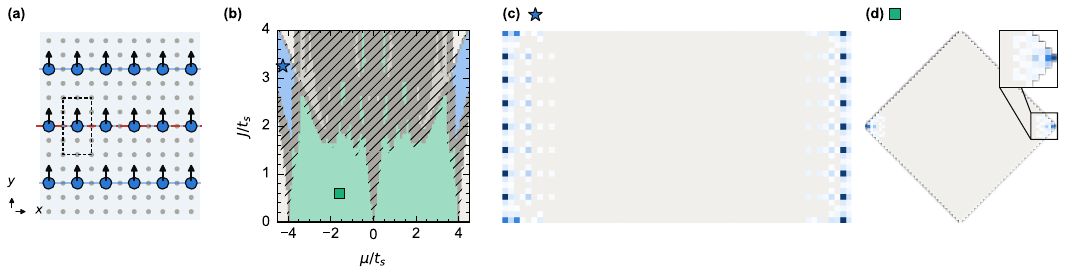}
    \caption{Magnetic-adatom superlattice: one classical spin (exchange $J\tau_0\sigma_y$) per $2\times4$
    supercell of a square substrate ($t_s=1$, $D_x=0.5$, $D_y=0.4$, $\alpha=0.2$). (a)~Geometry: substrate
    sites (gray), adatoms on the sites $x\equiv0\pmod 2$, $y\equiv0\pmod 4$ with in-plane moments along $y$,
    perpendicular to the adatom chains along $x$, and the mirror line $y\to-y$ (red dash-dotted) through an adatom row; dashed box: the $2\times4$ supercell.
    (b)~Phases from the
    mirror-resolved zero modes of a ribbon ($100$ columns, periodic along $y$), colors as in
    \cref{fig:phases}(a); hatched: bulk gap below $0.03$ or zero modes not resolved. The line $J=0$ is
    time-reversal symmetric. Star and square: the points $(\mu,J)=(-4.25,3.25)$ and $(-1.6,0.6)$ of (c) and (d). (c),(d)~Density
    $\sum_n|\psi_n(\bm r)|^2$ of the states with $|E|<0.01$, linear scale, normalized to its maximum:
    (c)~the flat band on a $60\times33$ rectangle, $y=-16,\dots,16$, whose $9$ adatom rows ($y=0\bmod4$,
    including both outer rows) each end in one Majorana at each $x$-normal edge, giving $2\times9=18$ edge levels below $1.4\times10^{-3}$ (next level $0.15$);
    (d)~the two zero modes of the double flip on the diamond $|x|+|y|\le32$ (next level $0.03$), with a
    zoom on the corner $(32,0)$.}
    \label{fig:shiba}
\end{figure*}

\section{Magnetic-adatom model}
\label{sec:adatoms}

\Cref{sec:model,sec:sectors,sec:realspace,sec:protection} use a minimal model chosen for its
symmetries rather than for its origin, and a model is useful only if it can be realized. Magnetic
adatoms on superconductors are a natural platform: they can be placed one by one, and their exchange
field breaks time reversal locally. In this
section we map the minimal model term by term onto a microscopic Hamiltonian, discuss which substrates
have a gap odd under the mirror, and place classical spins on a triplet substrate. The resulting adatom
superlattice realizes the double flip phase for a weak exchange and the flat band phase for a strong exchange near
the substrate band edge. In this section, we read both phases on flakes, estimate their energy scales and their
tolerance to imperfect moments, describe their spectroscopic signatures, and state the limits of the
model.

\Cref{eq:model} is, term by term, the Bogoliubov--de Gennes Hamiltonian of a spinful helical $p$-wave
superconductor~\cite{QiHughesRaghuZhang2009} with Rashba coupling in an in-plane exchange field. In the
Nambu basis $(c_\uparrow,c_\downarrow,c^\dagger_\uparrow,c^\dagger_\downarrow)$, the normal part
$h(\bk)=\epsilon_\bk\sigma_0+J\sigma_y+\alpha(\sin k_x\,\sigma_y-\sin k_y\,\sigma_x)$ and the pairing
$\Delta(\bk)=D_x\sin k_x\,\sigma_x+iD_y\sin k_y\,\sigma_0$ reproduce \cref{eq:model} exactly, with
$m_0=J$, $b_1=-\alpha$, $a=D_x$ and $b_2=-D_y$ (\cref{app:shiba}). Rashba coupling generates both $b_1$ terms with a common coefficient. At $\alpha=0$, the operator $\tau_x\sigma_z$
anticommutes with every other term, and the flat band would be the familiar chiral one.

The superconducting pairing potential must be odd under the mirror, as only then does $\Xi$ act within each mirror sector and make it a
class-D problem~\cite{UenoYamakageTanakaSato2013}. Conventional spin-singlet $s$-wave pairing is even under every mirror, so
Shiba lattices with such pairing~\cite{Yu1965, Shiba1968, Rusinov1969, NadjPerge2013,
PientkaGlazmanvonOppen2013, RontynenOjanen2015, LiNeupert2016, Soldini2023} cannot realize the
sector-resolved phases, while the helical $p$-wave pairing potential above is odd (\cref{app:shiba}). With an even pairing potential the mirror commutes with
$\Xi$ and exchanges the sectors; the mirror-protected phases of Ref.~\cite{Soldini2023} instead rest on
mirrors combined with time reversal, with winding numbers on the mirror lines. The flat band needs only the $p_x$ component of the pairing potential. In the $\mathbf d$-vector notation
$\Delta(\bk)=[\mathbf d(\bk)\cdot\boldsymbol\sigma]\,i\sigma_y$~\cite{SigristUeda1991}, this component
has $\mathbf d\parallel z$, so an exchange field along $y$ is not pair-breaking. The double flip phase also
needs the $p_y$ component, with $\mathbf d\parallel y$ (\cref{app:shiba}). One candidate substrate with a gap odd under a mirror is UTe$_2$~\cite{Ran2019, Aoki2022}, which has point group $D_{2h}$ but whose pairing symmetry is still debated.
A pairing state odd under $a\to-a$ is $B_{3u}$ or $A_u$, and one odd under $b\to-b$ is $B_{2u}$ or $A_u$. On the
$(0\bar11)$ cleave plane~\cite{Jiao2020}, only the former mirror survives, so chains along $[011]$ need
$B_{3u}$ or $A_u$ pairing. Whether a
two-dimensional model with a surface Rashba splitting below the triplet gap applies to that surface is an
open question, and we treat the following as a minimal model,
not a quantitative prediction.

Energies in the rest of this section are in units of the substrate hopping $t_s$. We place one classical spin with exchange $J$ on every $2\times4$ supercell of a square substrate with
triplet gaps $D_x$, $D_y$ and Rashba coupling $\alpha$ [\cref{fig:shiba}(a); values in the caption].
The moments lie in plane and perpendicular to the chains~\cite{BalatskyVekhterZhu2006,
Kaladzhyan2016}. Resolving the zero modes of
a ribbon by the mirror on both mirror lines gives the sector indices and the phase diagram of
\cref{fig:shiba}(b) (\cref{app:shiba}). The bare substrate ($J=0$) is a time-reversal-symmetric helical
superconductor, whose gapless helical edge covers the whole boundary of a flake (\cref{app:shiba}). A weak
exchange breaks time reversal and, by the in-plane-field mechanism of Refs.~\cite{Zhu2018, Pahomi2020},
turns it into the double-flip phase over most of the $(\mu,J)$ plane below
$J\approx2$, gapped up to $0.20$. On a diamond-oriented flake, the double-flip phase exhibits only two zero modes ($|E|\approx10^{-7}$, next
level $0.03$) that sit at the two mirror-axis corners, with $96\%$ of their weight within six sites of them
[\cref{fig:shiba}(d)]. Near the substrate band edges, a stronger exchange, $J\approx2.5$--$4$, drives the lattice into the
flat-band phase, with a gap up to $0.18$. Its zero modes line the $x$-normal edges of a rectangular
flake, one Majorana per adatom row at each edge [\cref{fig:shiba}(c)]. On a ribbon, their splitting
falls from $2\times10^{-11}$ to $10^{-14}$ between $100$ and $160$ columns, a decay length of about $8$
columns. Dense
chains, one adatom per substrate site along $x$ and three to four sites apart along $y$, widen
the flat-band window to about $2$ in $\mu$, raise the gap to $0.3$ and lower the onset to
$J\approx1.5$ (\cref{app:shiba}). The flat band phase thus needs an exchange of the order of the substrate hopping and
a chemical potential near the substrate band edge. This is where a fixed triplet gap is least justified. At
such an exchange, the adatoms act as strong magnetic scatterers rather
than as a dilute Shiba lattice, so the classical-spin, fixed-gap treatment is at best qualitative.

We also read the indices directly on adatom flakes, with adatom rows playing the role of the rows
in
\cref{eq:flakes}. Two flakes with an odd and an even number of adatom rows reproduce the ribbon phase at
$1500$ of $1636$ gapped grid points (\cref{app:shiba}), failing only where the localizer gap is small, mostly
next to the gapless helical line. We then add mirror-symmetric disorder, consisting of an on-site
potential uniform in $[-w,w]$, a relative spread of the adatom exchange uniform in $[-w/2,w/2]$, and
missing adatoms with probability $0.1$. At the double-flip point $(\mu,J)=(-1.6,0.6)$ the reading is unchanged in all five
samples at each of $w=0.25$, $0.5$ and $1$, the last about five times the bulk gap. The flat
band is read correctly in $4$ of $5$ samples at
$w=0.25$, $1$ of $5$ at $w=0.5$ and none at $w=1$. Because the flat-band phase occupies a window
narrower than $1$ in $\mu$ next to the band edge, potential disorder of the order of its gap ($0.17$)
moves parts of the flake out of the phase (\cref{app:shiba}).

The gaps of both phases stay near a quarter of the triplet gap $\Delta_{\max}=\max_\bk|\Delta(\bk)|$
down to $\Delta_{\max}=0.03$, about $80~\mu$eV for a $0.3$~meV triplet gap. This requires the Rashba
splitting, and for the double flip phase also the exchange, to scale with $\Delta_{\max}$. At fixed
$\alpha=0.2$, both phases close for small triplet gaps (\cref{app:shiba}). Magnetic-adatom lattices are assembled atom by atom~\cite{Kim2018, Schneider2021,
Soldini2023}, but the moments must stay in plane and perpendicular to the chains. Let $\theta$ be the angle by which the moments tilt out of the substrate plane, from
$\hat y$ toward $\hat z$, so that the exchange becomes $J(\cos\theta\,\tau_0\sigma_y+\sin\theta\,\tau_z\sigma_z)$.
This tilt, or a singlet admixture $\lambda_s\tau_y\sigma_y$, breaks $\A$. The edge band then widens linearly, to $0.016\,\theta$ ($\theta$ in radians) and
$0.04\,\lambda_s$. The corner modes of the double flip, protected by the edge gap, stay below $10^{-5}$
up to $\theta=30^\circ$ and $\lambda_s=0.1$ (\cref{app:shiba}). The tilt
splitting is thus $0.025\,\theta\Delta_{\max}$, with $\theta$ in radians and $\Delta_{\max}\approx0.64$ at
$D_x=0.5$, $D_y=0.4$. For $\Delta_{\max}=0.3$~meV, this is about $0.7~\mu$eV at $5^\circ$ and $4~\mu$eV at $30^\circ$,
below a spectroscopic resolution of order $10~\mu$eV. The signature specific to $\A$ is therefore
measurable only for an exchange as strong as in \cref{eq:model}. Spiral
order~\cite{BrauneckerSimon2013, Klinovaja2013, VazifehFranz2013} breaks $M_y$ outright. The model
treats the spins classically, does not compute the gap self-consistently, and does not project onto
Shiba bands; in particular, the effect of the exchange on the $D_y$ component is not
computed (\cref{app:shiba}).

In scanning tunneling spectroscopy, the flat band would appear as a zero-bias peak confined to the
$x$-normal edges of the adatom array. The peak would be absent on the $y$-normal edges and in the
gapped interior. Its conductance grows with the edge length and is reduced only gradually by disorder
(\cref{sec:protection}). The confinement to one pair of edges and the gapped interior distinguish it
from the Andreev states of nodal superconductors, whose bulk is gapless. The peak alone does not
distinguish protection by $\A$ from an accidental chiral symmetry, which would give the same signal. The signature specific to $\A$, the linear splitting under tilted moments, is below spectroscopic
resolution in this model. The double flip would appear as zero-bias peaks only at the two
corners on the mirror axis, with the rest of the boundary gapped [\cref{fig:shiba}(d)].

\section{Conclusion}
\label{sec:conclusion}

In this work, we have shown that a mirror-symmetric superconductor whose pairing potential is odd under the mirror
can host a Majorana flat band across the whole edge Brillouin zone, with a fully gapped bulk and without
chiral symmetry or time reversal. The protecting operation, the mirror combined with particle--hole
conjugation into an antisymmetry $\A$ squaring to $+1$, protects Majorana arcs in nodal
superconductors~\cite{ChiuSchnyder2014}. Once the bulk is gapped, these arcs close into a band.
 A minimal four-band model with exactly these symmetries has a
closed-form phase diagram, because its mirror sectors are Kitaev chains, and realizes both the weak
indices and the strong $\Z_2$ of its crystalline class~\cite{ShiozakiSato2014}. Its phases include a flat band, a
first-order phase with two Majoranas at each end of a wire, and an intrinsic second-order phase. A spectral localizer made real by particle--hole symmetry reads the weak index on finite flakes and the
strong index on mirror-symmetric ones. Resolved by the mirror symmetry, it gives the four sector indices and,
from open flakes alone, the whole phase diagram.

The flat band is protected only as long as translation along the edge is preserved; disorder reduces it
to a Majorana parity, although its weak index, read with the localizer, survives disorder of several
times the bulk gap. Its extensive degeneracy invites interaction effects that we have not
studied. The second-order phase is the more robust of the two against perturbations that break the mirror,
since its corner Majoranas are single modes. In the magnetic-adatom model, the flat
band requires a strong exchange near the substrate band edge, while the double flip phase appears already
for a weak exchange. Platforms in which a mirror-odd gap is engineered rather than intrinsic,
for instance through a $\pi$ phase difference imposed across the mirror plane, as in planar Josephson
junctions~\cite{PientkaEtAl2017, HellLeijnseFlensberg2017}, would avoid the need for
an unconventional triplet substrate.

Beyond the specific models we considered, these results establish three general points. First, a fully gapped
Majorana flat band does not require chiral symmetry. A superconducting gap that is odd under a mirror
suffices, as long as translation along the edge is preserved, and this criterion can be checked
directly for any candidate material or engineered lattice. Second, because the mirror sectors of the
model are exact Kitaev chains, every real-space index has a closed form. The model therefore serves as
a solvable benchmark for spectral-localizer methods built on antiunitary antisymmetries, against which
numerical readings can be checked exactly. Third, the model places a weak phase, a first-order
crystalline phase and a second-order phase of a single symmetry class side by side, and they respond
very differently to disorder. The flat band, whose protection relies on translation along the edge,
reduces to Majorana parities, whereas the single corner Majoranas of the second-order phase survive.
This distinction should apply to any flat band protected by a symmetry that acts within slices of fixed
edge momentum.
\vspace{0.15cm}
\begin{acknowledgments}
J.S.\ acknowledges discussion with Isidora Araya Day. This work was performed in part at the Center for Integrated Nanotechnologies, an Office of
Science User Facility operated for the U.S.\ Department of Energy (DOE) Office of Science.
Sandia National Laboratories is a multimission laboratory managed and operated by National
Technology \& Engineering Solutions of Sandia, LLC, a wholly owned subsidiary of Honeywell
International, Inc., for the U.S.\ DOE's National Nuclear Security Administration under
Contract No.\ DE-NA-0003525. The views expressed in the article do not necessarily represent
the views of the U.S.\ DOE or the United States Government.
\end{acknowledgments}

\normalem
\bibliography{tsc}

\setcounter{secnumdepth}{5}
\renewcommand{\theparagraph}{\bf \thesubsubsection.\arabic{paragraph}}

\renewcommand{\thefigure}{S\arabic{figure}}
\setcounter{figure}{0}
\renewcommand{\thetable}{S\arabic{table}}
\setcounter{table}{0}

\clearpage
\onecolumngrid
\begin{appendix}
    \section{Notation}
\Cref{tab:notation} lists the symbols used in the paper, with the
equation or section where each is defined.
\begin{table}[!h]
\centering
\begin{tabular}{p{0.2\linewidth}p{0.55\linewidth}p{0.17\linewidth}}
\hline\hline
\raggedright Symbol & \raggedright Meaning & \raggedright Defined in \tabularnewline
\hline
\raggedright $\Xi=\tau_x\K$ & \raggedright particle--hole operation, $\Xi^2=+1$ & \raggedright \cref{eq:factors} \tabularnewline
\hline
\raggedright $M_y$, $\Xi'$ & \raggedright spinful mirror about $y$; particle--hole operation with a $\pi$ shift of the pairing phase & \raggedright \cref{eq:factors} \tabularnewline
\hline
\raggedright $\A=M_y\Xi'$ & \raggedright antiunitary antisymmetry, $\A^2=+1$, acting within each slice of fixed $k_y$ & \raggedright \cref{eq:A} \tabularnewline
\hline
\raggedright $U_\A=\tau_y\sigma_y$ & \raggedright matrix part of $\A$, real and symmetric & \raggedright \cref{eq:sym} \tabularnewline
\hline
\raggedright $W=\Xi\A$ & \raggedright unitary reflection $k_y\to-k_y$, $W^2=-1$ & \raggedright \cref{eq:W} \tabularnewline
\hline
\raggedright $iW$ & \raggedright Hermitian reflection of class $\mathcal R_+$ of the crystalline classification & \raggedright \cref{app:invariants} \tabularnewline
\hline
\raggedright $s=\pm1$ & \raggedright $W$ sector, eigenvalue of $-iW$ & \raggedright \cref{eq:sectorH} \tabularnewline
\hline
\raggedright $\nu(k_y)$, $\nu$ & \raggedright Kitaev index of the slice $k_y$; its common value in a gapped bulk & \raggedright \cref{eq:nuk} \tabularnewline
\hline
\raggedright $\nu_s^{k_y}$ & \raggedright Majorana number of $W$ sector $s$ on the mirror line $k_y\in\{0,\pi\}$ & \raggedright \cref{eq:nusector} \tabularnewline
\hline
\raggedright $\chi=\nu_+^0\nu_+^\pi$ & \raggedright strong $\Z_2$ index, relative to a site-row mirror & \raggedright \cref{eq:chi} \tabularnewline
\hline
\raggedright $\nu_y(k_x)$ & \raggedright weak index of chains along $y$ & \raggedright \cref{app:invariants} \tabularnewline
\hline
\raggedright $L(x)$, $\kappa$ & \raggedright particle--hole spectral localizer and its scaling parameter & \raggedright \cref{eq:loc} \tabularnewline
\hline
\raggedright $\zeta(x)$, $\zeta_s$ & \raggedright Pfaffian sign of the localizer; of its restriction to $W$ sector $s$ & \raggedright \cref{eq:zeta}, \cref{app:localizer} \tabularnewline
\hline
\raggedright $\zeta_{\rm even}$, $\zeta_{\rm odd}$ & \raggedright $\zeta$ on flakes of even and odd width & \raggedright \cref{eq:flakes} \tabularnewline
\hline
\raggedright $\chi_c$ & \raggedright $\chi$ relative to the flake's own mirror line & \raggedright \cref{app:localizer} \tabularnewline
\hline
\raggedright $n_s$ & \raggedright number of Majoranas that a corner binds in $W$ sector $s$ & \raggedright \cref{app:corners} \tabularnewline
\hline
\raggedright $\beta$ & \raggedright wedge slope, wedge $|x|+|y|/\beta\le R$ with corners $(\pm R,0)$ & \raggedright \cref{app:corners} \tabularnewline
\hline
\raggedright $w$ & \raggedright disorder strength, potentials uniform in $[-w,w]$ & \raggedright \cref{app:transport} \tabularnewline
\hline
\raggedright $t_c$, $\gamma$ & \raggedright lead--flake hopping and the resulting broadening of edge levels & \raggedright \cref{app:transport} \tabularnewline
\hline
\raggedright $\Delta_{\max}$ & \raggedright $\max_\bk|\Delta(\bk)|$, the largest triplet gap & \raggedright \cref{app:transport} \tabularnewline
\hline\hline
\end{tabular}
\caption{Notation.}
\label{tab:notation}
\end{table}

\section{Model, parameters and symmetries}
\label[appendix]{app:model}

This appendix collects the parameter sets used throughout the paper and the symmetry analysis of
\cref{eq:model} behind \cref{sec:model}. We list the symmetry operators and the sixteen terms they allow.
By testing every subset of five to eight of these terms, we show that the eight-term model is minimal and generic, and we
identify the symmetry restored when any one term is removed. We then discuss the residual $C_{2z}\mathcal
T$ symmetry and its fragile topology, and compare the antisymmetry $\A$ with other mechanisms for
Majorana flat bands.

\subsection{Parameter sets}
All calculations use \cref{eq:model} with one of the coupling sets of \cref{tab:params}. The main
text fixes the couplings at $t=1$, $t_y=0.45$, $a=0.8$, $b_1=0.3$, $b_2=1$ and varies the chemical
potential $\mu$ and the exchange field $m_0$. We call these the main-text couplings; \cref{fig:phases} shows the $(\mu,m_0)$ plane at these couplings.
Most checks in \cref{app:invariants,app:localizer,app:transport} use a second set, with $t_y=0.15$ and
$b_2=0.13$, at two points. The reference point $(\mu,m_0)=(1.5,1.5)$ lies in the flat-band phase and the
reference $(-,-)$ point $(-0.5,0.3)$ in the $(-,-)$ phase. They play the roles of the flat-band and
$(-,-)$ points at the main-text couplings. The adatom lattice of \cref{app:shiba} has its own parameters
and points.

\begin{table}[h]
\centering
\begin{tabular}{llll}
\hline\hline
name & couplings & $(\mu,m_0)$ & phase, bulk gap \\
\hline
main-text couplings & $t=1$, $t_y=0.45$, $a=0.8$, $b_1=0.3$, $b_2=1$ & varied & --- \\
trivial point & $t=1$, $t_y=0.45$, $a=0.8$, $b_1=0.3$, $b_2=1$ & $(3.5,0.2)$ & trivial, $0.39$ \\
flat-band point & $t=1$, $t_y=0.45$, $a=0.8$, $b_1=0.3$, $b_2=1$ & $(2.4,2)$ & flat band, $0.36$ \\
$(-,-)$ point & $t=1$, $t_y=0.45$, $a=0.8$, $b_1=0.3$, $b_2=1$ & $(0.3,0.2)$ & $(-,-)$, $0.34$ \\
double-flip point & $t=1$, $t_y=0.45$, $a=0.8$, $b_1=0.3$, $b_2=1$ & $(-2,0.2)$ & double flip, $0.36$ \\
\hline
reference point & $t=1$, $t_y=0.15$, $a=0.8$, $b_1=0.3$, $b_2=0.13$ & $(1.5,1.5)$ & flat band, $0.49$ \\
reference $(-,-)$ point & $t=1$, $t_y=0.15$, $a=0.8$, $b_1=0.3$, $b_2=0.13$ & $(-0.5,0.3)$ & $(-,-)$, $0.38$ \\
\hline
diamond point (\cref{app:corners}) & $t=1.2$, $t_y=0.85$, $a=0.8$, $b_1=0.5$, $b_2=2$ & $(0.05,0.05)$ & $(-,-)$, $0.19$ \\
\hline\hline
\end{tabular}
\caption{Parameter points. Bulk gap: minimum of $|E|$ over a dense $\bk$ grid, refined by local
minimization from its lowest points.}
\label{tab:params}
\end{table}

\subsection{Symmetry operators and allowed terms}
With $H(\bk)$ the $4\times4$ Bloch Hamiltonian in the
basis $\tau\otimes\sigma$, particle--hole symmetry and the antisymmetry read
\begin{equation}
\label{eq:sym}
    \tau_xH(\bk)^*\tau_x=-H(-\bk),\qquad
    U_\A H(k_x,k_y)^*U_\A=-H(-k_x,k_y),\qquad U_\A=\tau_y\sigma_y,
\end{equation}
with $U_\A$ real, symmetric and $U_\A^2=\id$, so $\A^2=+1$. Their product is the unitary
reflection $W=\Xi\A$ with internal part $\tau_x(\tau_y\sigma_y)^*=i\tau_z\sigma_y$, so that
$WH(k_x,k_y)W^{-1}=H(k_x,-k_y)$ and $W^2=-\id$. Both antiunitaries
anticommute with $W$, $\Xi W\Xi^{-1}=-W$ and $\A W\A^{-1}=-W$.

We write on-site and nearest-neighbor terms as $f(\bk)\,\Gamma$, with $\Gamma=\tau_i\sigma_j$ and
$f\in\{1,\cos k_x,\cos k_y,\sin k_x,\sin k_y\}$. By \cref{eq:sym}, a term is allowed if
$\tau_x\Gamma^*\tau_x=\mp\Gamma$ for $f$ even/odd in $\bk$ and $U_\A\Gamma^*U_\A=\mp\Gamma$ for $f$
even/odd in $k_x$. Of the $80$ candidates, $16$ survive (\cref{tab:terms}). \Cref{eq:model} uses the
eight marked with $\sharp$.

\begin{table}[h]
\centering
\begin{tabular}{ll}
\hline\hline
$f(\bk)$ & allowed $\tau_i\sigma_j$ \\
\hline
$1$ & $\tau_z\sigma_0^\sharp$, $\tau_0\sigma_y^\sharp$ \\
$\cos k_x$ & $\tau_z\sigma_0^\sharp$, $\tau_0\sigma_y$ \\
$\cos k_y$ & $\tau_z\sigma_0^\sharp$, $\tau_0\sigma_y$ \\
$\sin k_x$ & $\tau_0\sigma_0$, $\tau_x\sigma_x^\sharp$, $\tau_x\sigma_z$, $\tau_y\sigma_x$, $\tau_y\sigma_z$, $\tau_z\sigma_y^\sharp$ \\
$\sin k_y$ & $\tau_0\sigma_x^\sharp$, $\tau_0\sigma_z$, $\tau_x\sigma_0$, $\tau_y\sigma_0^\sharp$ \\
\hline\hline
\end{tabular}
\caption{The sixteen terms allowed by $\Xi$ and $\A$; $\sharp$ marks the eight terms of \cref{eq:model}. $\cos k\,\tau_0\sigma_y$ is a spin-dependent hopping.}
\label{tab:terms}
\end{table}

\subsection{Census and genericity}
For every subset of $n$ allowed terms, we ask two questions. Test~(i) concerns generic symmetry. For
random coefficients, are the four operators $\id$, $W$, $\Xi$ and $\A$ the only $\bk$-independent
$4\times4$ operators that map $H$ to $\pm H$, unitarily or antiunitarily, under $\bk\to\pm\bk$,
$(-k_x,k_y)$ or $(k_x,-k_y)$? Test~(ii) concerns topology. Does some choice of coefficients, drawn
uniformly from $[-2,2]$, give $\nu=-1$ on both mirror lines with a bulk gap? We use $400$ draws, and
$3000$ for $n=6$. A coefficient choice counts as gapped if the bulk gap, the smallest $|E|$ of $H(\bk)$
on a $41\times41$ grid of the Brillouin zone, exceeds $0.05$ and, in addition, the zero-dimensional index $\sgn\pf[H(k_x^*,k_y)U_\A]$ is constant along both
lines $k_x^*\in\{0,\pi\}$, checked on $2001$ values of $k_y$. A sign change of this index forces a node
on that line, which a coarse grid can miss. The counts are
\begin{center}
\begin{tabular}{lccccc}
\hline\hline
$n$ & subsets & generic & generic, gapped, $\nu=-1$ & \ldots\ with full $\epsilon_\bk\tau_z$ & \ldots\ and no spin-dependent hopping\\
\hline
$5$ & $4368$ & $0$ & $0$ & $0$ & $0$ \\
$6$ & $8008$ & $1560$ & $0$ & $0$ & $0$ \\
$7$ & $11440$ & $5208$ & $437$ & $0$ & $0$ \\
$8$ & $12870$ & $8710$ & $1926$ & $216$ & $80$ \\
\hline\hline
\end{tabular}
\end{center}
With the grid criterion alone, the fourth column would read $960$ and $3471$ for $n=7$ and $8$, and the
fifth $240$. More than half of those models are therefore nodal on a line $k_x^*$. The last column is
unchanged. At $n=7$, the models with spin-independent hopping in both directions all lack the
chemical-potential term $\tau_z\sigma_0$. \Cref{eq:model} is one of the $80$ eight-term models with the
full dispersion and spin-independent hopping. The topology test is a random search, so a zero means
that no gapped $\nu=-1$ point was found, not a proof that none exists. Test~(i) does not include maps
that shift $\bk$ by half a reciprocal-lattice vector, so the ``generic'' counts are upper bounds. The
extended search below finds such a symmetry in every seven-term subset of \cref{eq:model}.

We also remove one term of \cref{eq:model} at a time and draw the remaining seven coefficients at
random. We then search for every $\bk$-independent $4\times4$ operator $U$, unitary or antiunitary, with
$UH(\bk)^{(*)}U^{-1}=\pm H(g\bk)$. Here $g\bk$ is any of $\bk$, $-\bk$, $(-k_x,k_y)$, $(k_x,-k_y)$,
$\bk+(\pi,0)$, $\bk+(0,\pi)$, $\bk+(\pi,\pi)$ and $-\bk+(\pi,\pi)$. The condition is linear in $U$ and is
imposed at twelve random $\bk$, and the dimension of its null space counts the solutions. Maps that
exchange $k_x$ and $k_y$ cannot be symmetries for $t\ne t_y$ and are not included. With all eight
terms, the only solutions are $\id$, $W$, $\Xi$ and $\A$, each once. Removing any one term adds the
operators of \cref{tab:necessity} and their products with $W$. Each is a single invertible product
$\tau_i\sigma_j$ up to a phase, and three independent random draws give the same result.
\begin{table}[h]
\centering
\footnotesize
\begin{tabular}{lll}
\hline\hline
term removed & additional operations & meaning \\
\hline
$m_0\,\tau_0\sigma_y$ & unitary antisymmetry $\tau_x\sigma_y$ at $\bk$; antiunitary $\tau_0\sigma_y$ at $-\bk$ & chiral and time reversal (DIII) \\
$\sin k_x\,\tau_z\sigma_y$ (Rashba) & unitary antisymmetry $\tau_x\sigma_z$ at $\bk$; antiunitary $\tau_0\sigma_z$ at $-\bk$ & chiral (BDI) \\
$\sin k_y\,\tau_0\sigma_x$ (Rashba) & antiunitary $\tau_z\sigma_z$ at $\bk$; unitary antisymmetry $\tau_y\sigma_z$ at $-\bk$ & $C_{2z}\mathcal T$ type \\
$b_2\sin k_y\,\tau_y\sigma_0$ & antiunitary $\tau_0\sigma_x$ at $\bk$; unitary antisymmetry $\tau_x\sigma_x$ at $-\bk$ & $C_{2z}\mathcal T$ \\
$a\sin k_x\,\tau_x\sigma_x$ & antiunitary $\tau_z\sigma_x$ at $\bk$; unitary antisymmetry $\tau_y\sigma_x$ at $-\bk$ & $C_{2z}\mathcal T$ type \\
$\mu\,\tau_z\sigma_0$ & unitary $\tau_x\sigma_y$ at $\bk+(\pi,\pi)$; antiunitary antisymmetry $\tau_0\sigma_y$ at $-\bk+(\pi,\pi)$ & sublattice \\
$-2t\cos k_x\,\tau_z\sigma_0$ & antiunitary antisymmetry $\tau_y\sigma_y$ at $\bk+(\pi,0)$ & sublattice \\
$-2t_y\cos k_y\,\tau_z\sigma_0$ & unitary $\tau_z\sigma_y$ at $\bk+(0,\pi)$ & sublattice \\
\hline\hline
\end{tabular}
\caption{Symmetries restored by removing one term of \cref{eq:model}. ``At $g\bk$'' means
$UH(\bk)^{(*)}U^{-1}=\pm H(g\bk)$; antisymmetries carry the minus sign.}
\label{tab:necessity}
\end{table}
The two Rashba terms are needed for different reasons. The $\sin k_x$ term removes the chiral
operation, and the $\sin k_y$ term removes a $C_{2z}\mathcal T$-type antiunitary. The terms $\mu$, $t$
and $t_y$ remove operations that combine an internal matrix with the sublattice sign $(-1)^{x+y}$,
$(-1)^x$ or $(-1)^y$. Qsymm does not test such operations, and test~(i) of the census does not include
them.

We repeated the symmetry search with Qsymm~\cite{Varjas2018}. The Hamiltonians are entered with
symbolic coefficients, so only symmetries valid for all couplings are reported. The candidates are all
elements of the square point group $C_{4v}$ combined with time reversal, particle--hole conjugation and
chiral operations. For \cref{eq:model}, Qsymm finds exactly four elements, the identity, $W$ (spatial
part $M_y$), $\Xi$ and $\A$ ($M_y$ combined with particle--hole conjugation). Their matrices equal ours
up to a phase, and $\A\A^*=\Xi\Xi^*=+\id$. With hopping, exchange field and $p_x$ pairing only, it finds
sixteen elements, including the chiral operation. Adding the Rashba term, which is equivalent to
\cref{eq:model} at $b_2=0$, leaves eight elements. The chiral operation is then absent, but
$C_{2z}\mathcal T$ [$(C_{2z}\mathcal T)(C_{2z}\mathcal T)^*=+\id$], $M_x\mathcal T$ and two unitary
antisymmetries with spatial parts $C_{2z}$ and $M_x$ remain, and the $p_y$ triplet $b_2$ removes all
four. At $m_0=0$, Qsymm finds time reversal with $\mathcal T\mathcal T^*=-\id$ and the chiral operation,
so the model is in class DIII. At $\mu=m_0$ it finds no element beyond the four, so the sublattice
structure of \cref{app:invariants} is not a symmetry. Qsymm cannot test maps such as
$\bk\to\bk+(\pi,\pi)$, which we checked directly. Finally, \texttt{bloch\_family} returns the sixteen
on-site and nearest-neighbor terms allowed by $\Xi$ and $\A$, as in \cref{tab:terms}.

\subsection{\texorpdfstring{Residual $C_{2z}\mathcal T$ and fragile topology}{Residual C2zT and fragile topology}}
Without $b_2$ the model keeps
$UH(\bk)^*U^{-1}=H(\bk)$ with $U=\tau_0\sigma_x$, a spinful
$C_{2z}\mathcal T$ squaring to $+1$, which is the symmetry used in Ref.~\cite{Lee2025}. Twelve of the sixteen allowed terms respect it. The four that break it are $\sin k_x\,\tau_x\sigma_z$,
$\sin k_x\,\tau_y\sigma_x$, $\sin k_y\,\tau_0\sigma_z$ and $\sin k_y\,\tau_y\sigma_0$ ($b_2$). With it, $H$ is real in the basis
$V=\tau_0\otimes\frac{1}{\sqrt2}\left(\begin{smallmatrix}1&i\\1&-i\end{smallmatrix}\right)$, and the two
negative-energy bands form a real bundle. This bundle has first Stiefel--Whitney
classes~\cite{AhnKimKimYang2018, AhnParkKimKimYang2019} $w_1$ along $x$ and $y$. When both vanish, it
also has an Euler class $e$, the winding of the $SO(2)$ Wilson loop along $k_x$ as $k_y$ goes around the
zone~\cite{AhnParkYang2019, BouhonBlackSchafferSlager2019}. An even nonzero $e$ would signal
fragile topology~\cite{PoWatanabeVishwanath2018, AhnParkYang2019}. In the flat-band phase $w_1$
along $x$ equals the Kitaev index and is nontrivial, so $e$ is undefined. We drew $50$ random models from the twelve compatible terms, with coefficients uniform in $[-2,2]$,
$w_1=0$ and a bulk gap above $0.03$ after minimization. The $42$ whose two mirror lines have the same
pattern of inverted $W$ sectors have $e=0$, and the $8$ with the pattern of the double flip have
$e=\pm1$. The value $|e|=2$ never occurs. The residual symmetry therefore adds no topology beyond the
sector indices. Its only invariant, $w_2=e\bmod2$, is stable and is odd exactly in the double flip. In \cref{eq:model} at $b_2=0$, every sample with $e\neq0$ is nodal. The term $b_2$, which opens the
double-flip gap, breaks $C_{2z}\mathcal T$.

\subsection{Relation to other flat-band mechanisms}
\Cref{tab:mechanisms} compares the protection of
the flat band with earlier mechanisms. The weak index of class D alone pins Majoranas only at the particle--hole-invariant edge momenta. The
mirror, through $\A=\Xi W$, extends this to every $k_y$. The same combined operation protects point
nodes away from the mirror plane and the Majorana arcs between their projections~\cite{ChiuSchnyder2014}.
In a fully gapped bulk, the arcs close into a band across the whole edge zone.
\begin{table}[h]
\centering
\footnotesize
\begin{tabular}{p{0.27\linewidth}p{0.22\linewidth}p{0.07\linewidth}p{0.15\linewidth}p{0.14\linewidth}}
\hline\hline
\raggedright system & \raggedright protecting symmetry & \raggedright bulk & \raggedright edge zero modes & \raggedright time reversal \tabularnewline
\hline
\raggedright $d$-wave and nodal superconductors~\cite{Hu1994, TanakaKashiwaya1995, SatoTanakaYadaYokoyama2011, SchnyderRyu2011} & \raggedright chiral & \raggedright nodal & \raggedright between projected nodes & \raggedright yes \tabularnewline
\hline
\raggedright reflection-symmetric nodal superconductors~\cite{ChiuSchnyder2014} & \raggedright reflection combined with $\Xi$ or chiral & \raggedright nodal & \raggedright between projected nodes & \raggedright --- \tabularnewline
\hline
\raggedright crystalline line-nodal superconductors~\cite{KobayashiSumitaYanaseSato2018} & \raggedright chiral & \raggedright nodal & \raggedright between projected nodes & \raggedright yes, or magnetic translation \tabularnewline
\hline
\raggedright weakly coupled chains~\cite{SenguptaZuticKwonYakovenkoDasSarma2001} & \raggedright chiral & \raggedright gapped & \raggedright whole edge zone & \raggedright yes \tabularnewline
\hline
\raggedright gapped, spin-rotation symmetric~\cite{LappLinkTimm2024} & \raggedright chiral and spin rotation & \raggedright gapped & \raggedright whole edge zone & \raggedright yes \tabularnewline
\hline
\raggedright weak class-D superconductors~\cite{SeroussiBergOreg2014} & \raggedright $\Xi$ & \raggedright gapped & \raggedright $k_y=0,\pi$ only & \raggedright no \tabularnewline
\hline
\raggedright lattices in inhomogeneous fields~\cite{SedlmayrAguiarHualdeBena2015} & \raggedright effective chiral & \raggedright gapped & \raggedright whole edge zone & \raggedright no \tabularnewline
\hline
\raggedright vortex lines~\cite{ZhangDoubleMajorana2026, ZhangMa2026} & \raggedright effective particle--hole~\cite{ZhangDoubleMajorana2026}; $k_z$-resolved Chern parity~\cite{ZhangMa2026} & \raggedright --- & \raggedright along the vortex line & \raggedright --- \tabularnewline
\hline
\raggedright this work & \raggedright $\A=\Xi W$ (mirror odd on the gap) & \raggedright gapped & \raggedright whole edge zone & \raggedright no \tabularnewline
\hline\hline
\end{tabular}
\caption{Mechanisms for Majorana flat bands. ``---'': depends on the realization.}
\label{tab:mechanisms}
\end{table}

\section{Momentum-space invariants}
\label[appendix]{app:invariants}

This appendix derives the momentum-space invariants of \cref{sec:symmetry,sec:sectors}. We show that
$\A$ makes every slice of fixed $k_y$ a class-D chain, whose Kitaev index is constant in a gapped bulk,
and that this protection needs $\A$ alone. We then reduce the model to its $W$ sectors, derive the phase
regions and whether each is gapped or nodal, relate the sector indices to the strong and weak indices,
and discuss the sublattice structure that appears at $\mu=m_0$.

\subsection{\texorpdfstring{The $\A$ index and the flat band}{The A index and the flat band}}
At fixed $k_y$, $h(k_x)\equiv H(k_x,k_y)$ is a one-dimensional Hamiltonian with
$U_\A h(k_x)^*U_\A=-h(-k_x)$ and $U_\A U_\A^*=+\id$. It is therefore in class D for every $k_y$, not only
on the mirror lines. At $k_x^*\in\{0,\pi\}$, $-k_x^*\equiv k_x^*$,
so $U_\A h^*=-hU_\A$, and with $h^T=h^*$ and $U_\A^T=U_\A$,
\begin{equation}
    \big(h(k_x^*)U_\A\big)^T=U_\A h(k_x^*)^*=-h(k_x^*)U_\A .
\end{equation}
$h(k_x^*)U_\A$ is therefore antisymmetric, with $\pf(hU_\A)^2=\det h$, and the Kitaev index of
the slice~\cite{Kitaev2001} is
\begin{equation}
\label{eq:nuk}
    \nu(k_y)=\sgn\left[\pf[H(\pi,k_y)U_\A]\,\pf[H(0,k_y)U_\A]\right].
\end{equation}
The basis is fixed by $U_\A$, so no gauge choice enters. Each Pfaffian is a real, continuous function of $k_y$ that vanishes only where the spectrum closes at
$k_x^*$. The index $\nu(k_y)$ is therefore constant along $k_y$ as long as the lines $k_x=0$ and $\pi$
stay gapped, and gap closings away from these lines do not change it. For \cref{eq:model},
\begin{equation}
\label{eq:pfclosed}
    \pf[H(k_x^*,k_y)U_\A]=m_0^2-\epsilon(k_x^*,k_y)^2+(b_1^2-b_2^2)\sin^2k_y,\qquad
    \epsilon=-2t\cos k_x^*-2t_y\cos k_y-\mu ,
\end{equation}
which gives $\nu$ in every region of \cref{fig:phases}(a).

On a ribbon open along $x$ and periodic along $y$, $k_y$ is conserved and $\A$
maps each slice to itself. If $\nu=-1$, every slice is a class-D chain with an odd Majorana number and binds one Majorana at each
end~\cite{Kitaev2001}. Since $\nu$ does not depend on $k_y$, these zero modes form a band pinned at $E=0$
across the edge Brillouin zone. Any perturbation that keeps $\A$ and translation along $y$ keeps the band
at zero. Breaking $\A$ with $\lambda\tau_y\sigma_y$, which keeps $\Xi$, gaps the band linearly. For
$\lambda\le0.2$, $|E|\simeq0.37\,\lambda|\sin k_y|$ at the flat-band point $(\mu,m_0)=(2.4,2)$ ($0.362$--$0.369$ over $k_y$) and $0.098\,\lambda|\sin k_y|$ at the reference point. Zeros remain
at $k_y=0,\pi$, where $\Xi$ alone makes each slice class D. The flat band does not rely on the minimal
form of \cref{eq:model}. We add the eight other allowed terms of \cref{tab:terms} with random coefficients
uniform in $[-0.15,0.15]$. At $(\mu,m_0)=(2.4,2)$ the bulk gap is then $0.494$, and the edge band of a
ribbon ($N_x=60$) is flat to $2.2\times10^{-12}$.

\subsection{\texorpdfstring{Protection by $\A$ alone}{Protection by A alone}}
At $b_2=0$, \cref{eq:model} keeps the $C_{2z}\mathcal T$ of
\cref{app:model}, so the flat band could in principle rely on it. Of the $80$ candidate terms, four keep $\Xi$ and $\A$ but break $C_{2z}\mathcal T$:
$\sin k_x\,\tau_x\sigma_z$, $\sin k_x\,\tau_y\sigma_x$, $\sin k_y\,\tau_0\sigma_z$ and $b_2$. Twelve keep $\Xi$
and $C_{2z}\mathcal T$ but break $\A$, among them an in-plane Zeeman field along $x$, $\tau_z\sigma_x$. At
$(\mu,m_0)=(2.4,2)$ with $b_2=0$ (bulk gap $0.365$), the width $\max_{k_y}\min|E|$ of the edge band of a
ribbon ($N_x=60$) is
\begin{center}
\begin{tabular}{lccc}
\hline\hline
added term & strength & bulk gap & width \\
\hline
none & --- & $0.365$ & $9\times10^{-11}$ \\
$\sin k_y\,\tau_0\sigma_z$ (keeps $\A$) & $0.1$, $0.3$, $0.6$ & $0.365$ & $(8$--$9)\times10^{-11}$ \\
$\sin k_x\,\tau_x\sigma_z$ (keeps $\A$) & $0.1$, $0.3$ & $0.369$, $0.402$ & $7\times10^{-11}$, $1\times10^{-11}$ \\
$\tau_z\sigma_x$ (keeps $C_{2z}\mathcal T$) & $0.01$, $0.05$, $0.1$ & $0.365$ & $1.5\times10^{-3}$, $7.4\times10^{-3}$, $1.5\times10^{-2}$ \\
\hline\hline
\end{tabular}
\end{center}
Breaking $C_{2z}\mathcal T$ leaves the band flat, while breaking $\A$ with $C_{2z}\mathcal T$ intact gaps
it linearly. The flat band is therefore protected by $\A$ alone. It then needs only the $p_x$ component
of the pairing, $\mathbf d=D_xk_x\hat z$. On a rectangular substrate with a polar axis along $z$ (point
group $C_{2v}$), $k_x\hat z$ and the $p_y$ component $k_y\hat y$ of $b_2$ belong to different irreducible
representations. The helical gap of \cref{eq:model} thus combines two of them, while the flat band needs
only one. With the spin--orbit term $\sin k_y\,\tau_0\sigma_z$ in place of $b_2$, the symmetry group is
again exactly $\{\id,W,\Xi,\A\}$, and removing any one of the eight terms restores a symmetry, as in
\cref{tab:necessity}. The double flip, in contrast, needs $b_2$. At $(\mu,m_0)=(-2,0.2)$ and $(-2,0.05)$
with $b_2=0$, with or without $\sin k_y\,\tau_0\sigma_z$ up to strength $1$, the bulk gap stays below
$0.03$, the resolution of a $101\times101$ grid. At $b_2=1$ it is $0.365$ and $0.398$.

\subsection{Mirror sectors}
On the lines $k_y\in\{0,\pi\}$, $[W,H]=0$ and
$H$ is block diagonal in the eigenspaces $-iW=s=\pm1$. Since $\Xi$ is antiunitary and anticommutes with $W$, $W\psi=i\psi$ implies
$W\Xi\psi=-\Xi(i\psi)=i\,\Xi\psi$. The operation $\Xi$ therefore maps each sector to itself, and so does
$\A$. Each sector is then a class-D chain with its own Majorana number $\nu_s$. In a real orthogonal
basis $O$ adapted to the sectors, $O^T(hU_\A)O$ is block diagonal, and $\pf(O^TMO)=\det O\,\pf(M)$
factorizes into sector Pfaffians. The factor $\det O$ cancels, so $\nu=\nu_+\nu_-$. In the eigenbasis of $\tau_z\sigma_y$ (the internal part of $-iW$), $\tau_z$ and
$\sigma_y=s\tau_z$ are diagonal, $\tau_x\sigma_x$ is off-diagonal, and the $\sin k_y$ terms vanish,
which gives the Kitaev chains of \cref{eq:sectorH} with $\xi_s=-2t\cos k_x-(\mu-sm_0+2t_y\cos k_y)$.
The Pfaffian of a Kitaev chain at $k_x^*$ is proportional to $\xi_s(k_x^*)$, so
$\nu_s=\sgn[\xi_s(0)\xi_s(\pi)]$, which is \cref{eq:nusector}. As checks, $\nu=\nu_+\nu_-$ holds at the reference point, at both $(-,-)$ points and at the trivial and double-flip points. The Wilson
loop~\cite{Zak1989, YuQiBernevigFangDai2011} of each sector's negative-energy band along $k_x$ gives a
hybrid Wannier center $1/2$ where $\nu_s=-1$ and $0$ where $\nu_s=+1$, to $10^{-6}$ at the reference
point, at both $(-,-)$ points and at the trivial point, as the equivalence of Pfaffian and Berry-phase invariants
requires~\cite{BudichArdonne2013}. Like any class-D index, $\nu_s$ is stable under adding trivial
bands. We measure them in real space on cylinders (\cref{app:localizer}).

\subsection{Phase regions: gapped or nodal}
Writing $c_s=\mu-sm_0$, \cref{eq:nusector} on the two mirror lines reads
$|c_s\pm2t_y|<2t$. For $t_y<t$, sector $s$ is inverted on both lines if $|c_s|<2t-2t_y$, on neither
if $|c_s|>2t+2t_y$, on $k_y=0$ only if $|c_s+2t|<2t_y$, and on $k_y=\pi$ only if $|c_s-2t|<2t_y$.
Combining the sectors, with $m_0\ge0$, gives \cref{tab:Sphases}. All boundaries lie on the lines $|\mu\mp m_0|=2t\pm2t_y$. On the $161\times121$ grid of \cref{fig:phases}, these regions agree with the numerically evaluated sector indices at every point not exactly
on a boundary.

\begin{table}[h]
\centering
\begin{tabular}{p{0.22\linewidth}p{0.42\linewidth}p{0.28\linewidth}}
\hline\hline
\raggedright phase & \raggedright region & \raggedright bulk \tabularnewline
\hline
\raggedright trivial & \raggedright $\big||\mu|-m_0\big|>2t+2t_y$ & \raggedright gapped \tabularnewline
\hline
\raggedright flat band ($\nu=-1$) & \raggedright $|\mu-sm_0|<2t-2t_y$ and $|\mu+sm_0|>2t+2t_y$ & \raggedright gapped \tabularnewline
\hline
\raggedright $(-,-)$ & \raggedright $|\mu|+m_0<2t-2t_y$ & \raggedright gapped \tabularnewline
\hline
\raggedright double flip & \raggedright $\big||\mu|-2t\big|+m_0<2t_y$ & \raggedright gapped for large $b_2$ \tabularnewline
\hline
\raggedright sector swap & \raggedright $|\mu|+|m_0-2t|<2t_y$ & \raggedright nodal (forced by the sector masses) \tabularnewline
\hline
\raggedright single flip & \raggedright otherwise & \raggedright nodal (forced by $\nu$) \tabularnewline
\hline\hline
\end{tabular}
\caption{Phases of \cref{eq:model} from the sector indices on the two mirror lines ($m_0\ge0$, $t_y<t$),
relative to the mirror about a site row. The sign patterns $(\nu_+^0,\nu_-^0,\nu_+^\pi,\nu_-^\pi)$ that
define each phase are listed in \cref{tab:phases}. The mirror about a bond row exchanges the
flat-band and sector-swap patterns.}
\label{tab:Sphases}
\end{table}

In a single flip $\nu(0)\ne\nu(\pi)$, so $\nu(k_y)$ must change and the
bulk must close on $k_x=0$ or $\pi$. In the sector-swap region and the double-flip phase, $\nu(0)=\nu(\pi)$, so $\nu$ forces no node. Whether the bulk is
gapped then depends on the couplings, with bulk gaps minimized from a $101\times101$ grid. At the swap
point $(\mu,m_0)=(0,2)$, the gap is below $10^{-15}$ for every $b_1\in\{0,0.1,0.3\}$ and
$b_2\in\{0.13,0.5,1\}$, and also at $t_y=0.35$. The node is forced. On the line $k_x=0$,
$\pf[H(0,k_y)U_\A]=m_0^2-(2t+\mu+2t_y\cos k_y)^2+(b_1^2-b_2^2)\sin^2k_y$ (\cref{eq:pfclosed}) is negative
at $k_y=0$ and positive at $k_y=\pi$ whenever $|\mu|+|m_0-2t|<2t_y$, for every $b_1$ and $b_2$. Its sign is
a zero-dimensional class-D index along the whole line, so it must vanish in between, and the same holds
on $k_x=\pi$. At $(0,2)$ and $(0.3,2.2)$, the signs at $(0,0)$, $(0,\pi)$, $(\pi,0)$, $(\pi,\pi)$ are
$(-,+,+,-)$. The argument uses the sector masses of \cref{eq:model}. The adatom lattice, whose sector
masses differ, has a gapped swap without Rashba coupling (\cref{app:shiba}). \Cref{tab:dfgap} gives the bulk gap at $(\mu,m_0)=(-2,0.05)$, in the double-flip phase. It shows that $b_2$ opens the gap
once it exceeds the Rashba tilt. At the main-text couplings ($b_2=1$), the double-flip phase is gapped
over $92\%$ of its area, with gaps up to $0.46$.

\begin{table}[h]
\centering
\begin{tabular}{lccc}
\hline\hline
 & $b_2=0.13$ & $b_2=0.5$ & $b_2=1$ \\
\hline
$b_1=0$ & $0.080$ & $0.25$ & $0.25$ \\
$b_1=0.3$ & $3\times10^{-16}$ & $0.19$ & $0.20$ \\
\hline\hline
\end{tabular}
\caption{Bulk gap at $(\mu,m_0)=(-2,0.05)$, in the double-flip phase, for $t=1$, $t_y=0.15$, $a=0.8$ and
the listed $b_1$, $b_2$.}
\label{tab:dfgap}
\end{table}

\subsection{Strong and weak indices}
Where the bulk is gapped, $\nu(0)=\nu(\pi)$, so
$\nu_+^0\nu_+^\pi=\nu_-^0\nu_-^\pi$. This sign is $-1$ only in the double flip and the sector
swap. For $\nu=+1$ it is the strong $\Z_2$ of class D with a reflection $iW$ that squares to $+1$ and
commutes with $\Xi$. This is the entry $\mathcal R_+$, $\Z_2$ in Table~I of Ref.~\cite{Langbehn2017}. See
also Ref.~\cite{ChiuYaoRyu2013}. For $\nu=-1$ the sign depends on the mirror line. The mirror about a bond
row differs from that about a site row by one lattice translation. This translation multiplies the $W$
eigenvalues at $k_y=\pi$ by $-1$, exchanges $\nu_\pm^\pi$, and maps the sign to its product with $\nu$
(compare the even-$L_y$ cylinders of \cref{eq:cyl}). The flat band and the sector swap differ only in
this sign. The index $\nu$ and the sector indices on one line are weak indices of chains stacked along
$y$. \Cref{app:corners} shows that $\tfrac12[1-\nu_+^0\nu_+^\pi]$ is the parity of the number of
Majoranas at a mirror-axis corner.

Chains along $y$ (slices of fixed $k_x$) carry the class-D index
$\nu_y(k_x)=\sgn\{\pf[H(k_x,0)\tau_x]\,\pf[H(k_x,\pi)\tau_x]\}$ at $k_x\in\{0,\pi\}$, defined by $\Xi$
alone. For \cref{eq:model}, at $k_x^*\in\{0,\pi\}$,
$\pf[H(k_x^*,k_y)\tau_x]=m_0^2-\epsilon(k_x^*,k_y)^2+b_1^2\sin^2k_y$. At $k_y\in\{0,\pi\}$ this is the
product of the two sector masses, so $\nu_y(k_x^*)=-1$ exactly when one sector changes sign between
$(k_x^*,0)$ and $(k_x^*,\pi)$. More generally, at the four particle--hole-invariant momenta
$(k_x^*,k_y^*)$ both Pfaffians square to $\det H$, so they agree up to sign. We find that they agree,
$\pf(HU_\A)=\pf(H\tau_x)$. For \cref{eq:model} the $\sin^2k_y$ terms vanish there, and for $500$ random
combinations of all sixteen allowed terms we found no exception, although we have no general proof. For
these models, $\nu_y(k_x^*)$ is therefore the product of the zero-dimensional $\A$ indices at the two ends
of the line $k_x^*$. This product is $+1$ whenever that line is gapped, so $\nu_y=+1$ in every gapped
phase. On a $401\times301$ grid over the $(\mu,m_0)$ plane at the main-text couplings, $\nu_y=-1$ occurs only in the
sector swap and in the single flip, both nodal. The product of the four Pfaffian signs, the parity of the
Chern number, is $+1$ in every gapped phase, as the mirror requires. It is $-1$ only in the single flip,
where $\nu(0)\ne\nu(\pi)$ forces the bulk to close.

\subsection{Sublattice structure}
The model has a sublattice structure that is not a symmetry but produces exactly degenerate zero
modes on some flakes. We describe it here so that these modes are not mistaken for topological ones. On
a mirror line, the sector chain of \cref{eq:sectorH} anticommutes with the sublattice sign $(-1)^x$,
$h_s(k_x+\pi)=-h_s(k_x)$, exactly when its effective chemical potential vanishes,
$\mu-sm_0+2t_y\cos k_y=0$. Both $\cos k_x$ and the Rashba tilt $\sin k_x$ then change sign under
$k_x\to k_x+\pi$. The sector is a chiral (BDI) chain with winding number $\pm1$, deep inside its inverted
region, so the class-D index is unchanged. For $t_y\to0$ the condition is $\mu=\pm m_0$. At finite $t_y$
it holds on one mirror line, at $\mu=sm_0-2t_y$ ($k_y=0$) or $\mu=sm_0+2t_y$ ($k_y=\pi$). We checked
numerically that no operator of the form $(-1)^{x+y}\otimes\tau_i\sigma_j$ anticommutes with the
Hamiltonian of a $12\times9$ flake restricted to either $W$ sector, at $\mu=m_0$ or away from it. The
property therefore does not survive as a two-dimensional symmetry of a sector.

The terms of \cref{eq:model} that commute with $\tau_z\sigma_y$ form,
in the block $\tau_z\sigma_y=s$, $h_s(\bk)=[-2t\cos k_x-2t_y\cos k_y-(\mu-sm_0)]\rho_z+a\sin
k_x\,\rho_y-s\,b_1\sin k_x\,\rho_0$. At $\mu=sm_0$ this block satisfies $h_s(\bk+(\pi,\pi))=-h_s(\bk)$, a sublattice chiral symmetry. The
terms $b_1\sin k_y\,\sigma_x$ and $b_2\sin k_y\,\tau_y$ couple the blocks and break it, so it is not a
symmetry of $H$. On the mirror lines, where these couplings vanish, it maps sector $s$ at $k_y=0$ onto
sector $s$ at $k_y=\pi$, $h_s(k_x+\pi,\pi)=-h_s(k_x,0)$ (to $4\times10^{-16}$). At $\mu=sm_0$, sector $s$ is
therefore inverted on both lines or on neither. The reference point ($\mu=m_0=1.5$) lies on the line
$\mu=m_0$, but its flat band does not rely on this structure. At $(\mu,m_0)=(1.2,1.5)$, $(1.8,1.5)$ and
$(2.0,1.2)$, with bulk gaps $0.39$, $0.47$ and $0.41$, the ribbon band ($N_x=60$) is flat to
$\max_{k_y}\min|E|=(0.9$--$1.1)\times10^{-10}$, as at the reference point. On flakes at $\mu=m_0$ the
structure produces exact zero modes, which split once $\mu\ne m_0$ (\cref{app:corners}). The points of
\cref{fig:densities} avoid it.

\section{Spectral localizer}
\label[appendix]{app:localizer}

This appendix gives the details of the real-space indices of \cref{sec:realspace}. We explain what the
full and sector localizers count, derive the sector indices on cylinders and the strip formula behind
\cref{eq:flakes}, and compare the resulting localizer phase map with the Bloch phase diagram. We then
show how the terminations affect the readings and close with practical remarks on the localizer gap.

\subsection{Localizers and what they count}
On a flake symmetric under $y\to-y$, $\A=O\K$ with $O=P\otimes\tau_y\sigma_y$ and $P$ the site
permutation $y\to-y$. The matrix $O$ is real, symmetric and orthogonal, $O^2=\id$, and
\begin{equation}
    OX^*O=+X,\qquad OY^*O=-Y,\qquad OH^*O=-H.
\end{equation}
The matrix $V=(O+i\id)/\sqrt2$ is unitary. For any Hermitian $Q$ with $OQ^*O=\pm Q$, the transformed
$Q_V=VQV^\dagger$ satisfies $Q_V^*=\pm Q_V$~\cite{Lee2025}. Hence $X_V$ is real symmetric, while $Y_V$ and
$H_V$ are $i$ times real antisymmetric matrices. In the $\A$-based localizer
\begin{equation}
\label{eq:locA}
    L_\A(x) = \kappa(X_V-x)\otimes\sigma_y+\kappa Y_V\otimes\sigma_x
    +H_V\otimes\sigma_z ,
\end{equation}
$\sigma_y$ is imaginary antisymmetric and $\sigma_x,\sigma_z$ are real symmetric. Every term is
therefore $i$ times a real antisymmetric matrix, and $B(x)=-iL_\A(x)$ is real and antisymmetric, just like
the particle--hole localizer of \cref{eq:loc}. Its eigenvalues come in pairs $\pm i\lambda$, so $\pf(B)$ is
real. It changes sign exactly when a pair passes through zero, that is, when the localizer gap
$g_L(x)=\min|\mathrm{spec}\,L(x)|$ closes. We define $\zeta(x)=\sgn\pf[B(x)]\,\sgn\pf[B(x_0)]$, relative to a
point $x_0$ far outside the flake, as in \cref{eq:zeta}. Since $H$ enters without an energy shift, $\zeta$ is a marker
at $E=0$.

For a one-dimensional class-D system with a real structure, $\kappa(X-x)\otimes\sigma_y+H\otimes\sigma_z$
is the odd-dimensional localizer of Refs.~\cite{LoringSchulzBaldes2017, DollSchulzBaldes2021}. For $x$
inside the sample and $\kappa$ small compared with the gap, $\zeta(x)$ is the $\Z_2$ index of the chain,
its Majorana number. The sign $\zeta$ cannot change with $\kappa$ or $x$ without closing the localizer
gap. We therefore regard a reading as reliable when the localizer gap is open, and check that it does
not change with $\kappa$. The localizer remains defined with disorder while its gap is
open~\cite{Stoiber2025}.

Neither $\A$ nor the $Y$ term is needed for $\zeta$. In the Majorana basis of $\Xi$ alone, $H_M$ is
imaginary antisymmetric and $X$ real, and the odd-dimensional localizer of \cref{eq:loc} measures the same
Majorana parity at an $x$ end. On $20\times8$ and $20\times9$ flakes ($\kappa=0.15$), it gives
$(\zeta_{\rm even},\zeta_{\rm odd})=(+,+)$, $(+,-)$, $(+,+)$ and $(-,-)$ at the trivial, flat-band,
$(-,-)$ and double-flip points of \cref{fig:densities}, as \cref{eq:flakes} requires. The localizer
gaps are $0.19$--$0.56$. We then add on-site disorder that breaks the mirror, random $\tau_z\sigma_0$ and
$\tau_z\sigma_z$ terms uniform in $[-w,w]$. The readings at the flat-band and double-flip points are
unchanged for $w=0.2$ and $0.4$, with localizer gaps $0.18$--$0.45$, although the $\A$-based construction
above is not defined there.

\subsection{Sector localizer and cylinders}
On the flake, $W=P\otimes i\tau_z\sigma_y$ commutes with $H$ and $X$ and anticommutes with $Y$. In the
$V$ basis, $\A$ is complex conjugation. Since $\A$ maps each eigenspace of $W$ to itself, each eigenspace
has a real orthonormal basis $P_s$. The operator $Y$ maps the eigenspace $W=+i$ to $W=-i$, so
$P_s^\dagger YP_s=0$ and
\begin{equation}
    -iL_s(x)=-i\left[\kappa(P_s^TX_VP_s-x)\otimes\sigma_y+P_s^\dagger H_VP_s\otimes\sigma_z\right]
\end{equation}
is real antisymmetric with no $Y$ term, by symmetry rather than by geometry~\cite{CerjanLoringSchulzBaldes2024}.
The sign $\zeta_s$ is the Majorana number of sector $s$ of the flake viewed as a wire along $x$, and
$\zeta_+\zeta_-$ is that of the whole wire. The $\A$-based localizer of \cref{eq:locA}, with its $Y$
term, gives the same sign, $\zeta=\zeta_+\zeta_-$, on every flake we tested. These are twelve flakes of
\cref{eq:model} at the trivial, flat-band, $(-,-)$ and double-flip points of \cref{tab:params} with $L_y=9$--$11$, and eighteen
adatom flakes (\cref{app:shiba}, with $\alpha=0$ and $0.2$), with localizer gaps $0.05$--$0.58$. In
particular, $\zeta$ is not $\nu$ itself. In the double-flip phase $\nu=+1$, but $\zeta=-1$ on every
width.

On a cylinder, open along $x$ and periodic along
$y$ with circumference $L_y$, transverse momentum $k_y=2\pi n/L_y$ is conserved. The mirror of the
flake, $y\to L_y-1-y$, passes through a site for odd $L_y$ and between two sites for even $L_y$.
The slices $k_y=0$ and, for even $L_y$, $k_y=\pi$ are mapped to themselves. In sector $s$ they
contribute their sector Majorana numbers, $\nu_s^0$ and $\nu_{-s}^\pi$. The index at $\pi$ belongs to the
other sector because a bond-centered mirror differs from the site-centered one by a translation, which
multiplies its eigenvalues at $k_y=\pi$ by $-1$ and exchanges the sectors there. Every other slice pairs
with $-k_y$. The mirror $W$ maps one partner onto the other, so each sector holds one copy of the slice,
on which $\A$ acts as a class-D particle--hole symmetry with index $\nu$. With $(L_y-1)/2$ pairs for odd and $(L_y-2)/2$ for even $L_y$,
\begin{equation}
\label{eq:cyl}
    \zeta_s=\begin{cases}\nu_s^0, & L_y\equiv1\\ \nu\,\nu_s^0, & L_y\equiv3\\
    \nu_s^0\,\nu_{-s}^\pi, & L_y\equiv2\\ \nu\,\nu_s^0\,\nu_{-s}^\pi, & L_y\equiv0\end{cases}\pmod4 .
\end{equation}
Two cylinders with $L_y\equiv1$ and $2\pmod4$ therefore give all four sector indices. We scanned a $21\times16$ grid
over the $(\mu,m_0)$ plane at the main-text couplings, with $L_y=9$ and $10$. All $132$ bulk-gapped points with
localizer gap above $0.05$ agree with \cref{eq:cyl} in both sectors and both widths. The $13$ points that
disagree have localizer gaps of $0.003$--$0.031$ and lie next to phase boundaries, where a circumference
of ten sites does not resolve the transverse momenta.

\subsection{Flakes: strip formula and phase map}
An $L_x\times L_y$ flake is a wire along $x$ whose ends are segments of the $x$-normal edge of length
$L_y$, bounded by the gapped $y$-normal edges. The parity of the number of Majoranas at one end has two
sources. The weak index contributes a factor $\nu$ per row, $\nu^{L_y}$ in all, which is how a weak
index shows up for $L_y$ stacked chains~\cite{FulgaPikulinLoring2016}. Each $W$-pinned crossing of the
$x$-normal edge, confined between two mirror-related gapped ends, contributes one more Majorana. The
number of such crossings is odd exactly when $\nu_+^0\nu_+^\pi=-1$, so
\begin{equation}
\label{eq:strip}
    \zeta_+\zeta_-=\nu^{L_y}\,\nu_+^0\,\nu_+^\pi .
\end{equation}
Even widths read $\chi=\nu_+^0\nu_+^\pi$ and odd widths $\nu\chi$, which is \cref{eq:flakes} of the main
text. This counting argument is heuristic, but \cref{eq:strip} is exact for long flakes. Away from its
$x$ ends, a flake is a strip with translation along $x$, and $\zeta$ is the Kitaev index of that strip,
$\sgn[\pf(-iH_M(0))\,\pf(-iH_M(\pi))]$, with $H_M(k_x)$ the strip Hamiltonian in the Majorana basis of
$\Xi$. Equivalently, $\zeta=\nu\chi_c$, with $\chi_c$ taken
relative to the flake's own mirror line, a site row for odd and a bond row for even $L_y$. The chain-end count of \cref{sec:realspace} gives the same signs. The cylinder formula \cref{eq:cyl} is derived independently.

In the decoupled limit $b_1=b_2=0$, the signs can be counted exactly. The $L_y$ transverse standing waves
are even or odd under $y\to-y$. The sector $W=+i$ collects the even waves of the internal sector
$\tau_z\sigma_y=+1$ and the odd waves of $\tau_z\sigma_y=-1$, and $\zeta_\pm=-1$ when the number of
inverted waves it collects is odd. At the reference point this gives $(\zeta_+,\zeta_-)=(-,+)$, $(-,-)$,
$(+,-)$ and $(+,+)$ for $L_y\equiv1,2,3,0\pmod4$. At the reference $(-,-)$ point it gives $(-,-)$ for odd and
$(+,+)$ for even $L_y$. At the double-flip point exactly one of the two signs is $-1$. Which sector carries the $-1$ is not a bulk property. Edge decoration flips both signs, and in the
double-flip diamond at $\mu=+2t$ the signs alternate between $(+,-)$ and $(-,+)$ with $L_y$. Only the
product is a bulk index.

\begin{figure}[h]
    \centering
    \includegraphics[width=\textwidth]{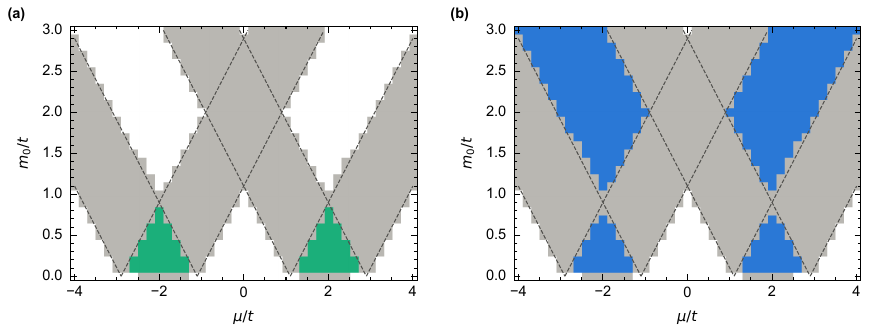}
    \caption{Localizer phase map over the $(\mu,m_0)$ plane at the main-text couplings: $\zeta_+\zeta_-$ at the center of
    $24\times L_y$ flakes, $\kappa=0.15\,t$ per lattice spacing; energies and gaps in units of $t$. Colored: $\zeta_+\zeta_-=-1$; white: $+1$; gray: bulk nodal (bulk gap
    below $0.03$, sector swap or single flip) or localizer gap below $0.02$. Dashed: the analytic
    boundaries $|\mu\mp m_0|=2t\pm2t_y$. (a)~$L_y=10$, where $\zeta_+\zeta_-=\nu_+^0\nu_+^\pi$ and the
    colored regions are the double-flip diamonds. (b)~$L_y=11$, where $\zeta_+\zeta_-=\nu\,\nu_+^0\nu_+^\pi$
    and the flat-band phase is added.}
    \label{fig:locmap}
\end{figure}

We evaluated $(\zeta_+,\zeta_-)$ on $24\times10$ and $24\times11$ flakes ($\kappa=0.15$) on a
$41\times31$ grid over the $(\mu,m_0)$ plane at the main-text couplings [\cref{fig:locmap}]. At all $507$ bulk-gapped
grid points with $m_0>0$, the product agrees with \cref{eq:strip} for both widths. Of these points, $469$
have a localizer gap above $0.05$ and $10$ have one below $0.02$. The smallest is $0.002$, and inside the
double-flip diamonds it is at least $0.02$. On the line $m_0=0$, the model has time reversal
$i\sigma_y\K$, squaring to $-1$, and is a helical class-DIII superconductor with gapless helical edges.
The localizer gap closes there, and only $23$ of $41$ points agree. The exchange field $m_0$ is what turns
the helical phase into the double flip~\cite{Zhu2018, Pahomi2020}. \Cref{fig:localizer}(a) refines
this map to an $81\times61$ grid and resolves the sectors to separate $(-,-)$ from the trivial phase. It
leaves unread the points where a second pair of flakes, $24\times14$ and $24\times15$, gives different
signs.

\subsection{Terminations}
That $\zeta$ is stable against weak disorder does not make every reading robust. Without the mirror,
the strong index read by $\zeta_{\rm even}$ is not protected against changes of the termination. To show
this, we attach one extra row along the top $y$-normal edge of $20\times L_y$ flakes. The row is a single
Kitaev chain given by \cref{eq:model} at $(\mu,m_0)=(2,1)$, where one sector is inverted, coupled to the
flake with $0.1$ times the nearest-neighbor $y$ hopping. It adds one Majorana at each $x$ end and flips
$\zeta$ at the trivial, flat-band and double-flip points of \cref{tab:params} for $L_y=8$ and $9$. The localizer gaps are
unchanged ($0.54$--$0.56$, $0.46$ and $0.19$). Decorating both edges symmetrically restores the readings.
The product $\nu=\zeta_{\rm odd}\zeta_{\rm even}$ of two identically terminated flakes is unaffected and
needs no mirror. In contrast, $\chi$ requires mirror-symmetric terminations, and the mirror enters the
sector localizers $\zeta_s$.

\subsection{Practical notes}
We close with two practical remarks on reading the localizer. The localizer gap is set by the smallest gap of the flake, bulk or edge. In the double flip this is the
gap of the $y$-normal edge, $0.19$ at the double-flip point, not the bulk gap. Near a phase boundary this
edge gap is small, and $\kappa$ must be reduced. At $(\mu,m_0)=(-2,0.05)$ with $t_y=0.15$ and $b_2=0.5$,
the edge gap is $0.039$, the localizer gap is $0.04$, and $\kappa=0.6$ fails. Wedge-shaped flakes are not
a reliable substitute for rectangles. In the flat-band phase their oblique edges carry a nearly flat
band, and the localizer gap closes (below $0.005$), as it can in gapless systems~\cite{FrancaGrushin2024}.
The index is then meaningless.

\section{Boundary and corner modes}
\label[appendix]{app:corners}

This appendix supports the discussion of boundary modes in \cref{sec:mm,sec:df}. We first count the
Majoranas on each edge. We then contrast the extrinsic corner modes of the $(-,-)$ phase, which depend on
the corner angle, with the intrinsic corner Majoranas of the double flip. Finally, we show that corner
modes can also appear on diamond flakes without a topological reason.

\subsection{Edges and counting}
Only the $x$-normal edges are mapped to themselves by $W$. Every other edge is gapped. At the reference
$(-,-)$ point, the gap is $0.38$ on the $y$-normal edge and $0.016$, $0.031$, $0.019$ and $0.011$ on edges along
$(2,1)$, $(1,1)$, $(1,2)$ and $(1,4)$. At the double-flip point, it is $0.19$ on the $y$-normal edge and
$0.05$--$0.14$ on the oblique ones. A flake with oblique edges therefore has a gapped boundary except
where the mirror axis meets it, and the question is what those points bind. We use wedges
$|x|+|y|/\beta\le R$, whose corners $(\pm R,0)$ lie on the mirror axis, and resolve the lowest states by $W$
sector and by corner.

Each $W$ sector carries its own particle--hole symmetry, so the number $n_s$ of Majoranas that a corner
binds in sector $s$ is defined mod $2$. A local perturbation that respects $\Xi$ and $W$ changes it only by
two. Decorating the two mirror-related edges with Kitaev chains adds one Majorana and its mirror image at
the corner, which recombine into one mode per sector, $(n_+,n_-)\to(n_++1,n_-+1)$. Corner Majorana
numbers $(1,1)$ are therefore robust against every local perturbation at the corner but can be removed by
changing the edges, as in an extrinsic or boundary-obstructed phase~\cite{Geier2018,
TrifunovicBrouwer2019, Khalaf2021}. Only the parity $n_++n_-$ is a bulk property. In the edge theory of
the $x$-normal edge, each $W$-pinned crossing, at $k_y=0$ and at $k_y=\pi$, acquires a mass on an oblique
edge that is odd under $W$. This mass changes sign at the mirror axis, so the crossing binds one Majorana
there~\cite{Langbehn2017}. For any gapped phase with $\nu=+1$,
\begin{equation}
    n_++n_- \equiv \tfrac12\left[1-\nu_+^0\,\nu_+^\pi\right] \pmod 2 .
\end{equation}

\subsection{\texorpdfstring{The $(-,-)$ phase: extrinsic corners}{The (-,-) phase: extrinsic corners}}
Both mirror lines carry a pinned crossing, so the parity is even, and the corner Majorana numbers depend
on the corner angle. \Cref{tab:corners} summarizes the corner levels. For $\beta=1/2$ and $\beta=1$, each corner
binds two Majoranas in the same sector, and a $\Xi$- and $W$-symmetric random perturbation within three
sites of the corner splits them linearly. For $\beta=2$, each corner binds one Majorana in each sector,
$(1,1)$, and neither that perturbation nor mirror-symmetric on-site disorder up to $w=1$ moves them. The
same holds at the main-text couplings [\cref{fig:corners}(c)]. These modes are pinned by $W$, since a uniform
$\lambda\tau_y\sigma_y$ splits them linearly, by $4\times10^{-4}\lambda$. Rectangles show the same
mechanism, with the $x$-normal edge in the role of the corner. A $40\times L_y$ flake has $4L_y$ zero modes
for $b_1=b_2=0$. With the interwire terms, it has exactly four for odd $L_y$ ($5$ to $21$) and none for
even $L_y$. For even $L_y$, the lowest level ($0.017$ at $L_y=8$, $0.011$ at $L_y=12$) does not depend on
$L_x$ and is the finite-size gap of the edge band. For odd $L_y$, the fourth-lowest $|E|$ at $L_y=15$ is
$9\times10^{-8}$, $6\times10^{-11}$ and $2\times10^{-14}$ for $L_x=40$, $60$ and $80$. The $(-,-)$ phase is
thus a first-order crystalline phase with $W$-pinned edge crossings, whose corner modes, where present,
are extrinsic.

\begin{table}[h]
\centering
\begin{tabular}{p{0.24\linewidth}p{0.08\linewidth}p{0.18\linewidth}p{0.2\linewidth}p{0.22\linewidth}}
\hline\hline
\raggedright point & \raggedright $\beta$ & \raggedright corner Majoranas & \raggedright lowest levels & \raggedright symmetric corner perturbation \tabularnewline
\hline
\raggedright $(\mu,m_0)=(-0.5,0.3)$ & \raggedright $1/2$, $1$ & \raggedright two, same sector & \raggedright split by up to $1.6\times10^{-4}$ and $3\times10^{-8}$, independent of $R$ & \raggedright linear, $(2$--$4)\times10^{-3}$ at strength $0.1$ \tabularnewline
\hline
\raggedright $(\mu,m_0)=(-0.5,0.3)$ & \raggedright $2$ & \raggedright $(1,1)$ & \raggedright $<4\times10^{-15}$ ($R=40$) & \raggedright $<3\times10^{-14}$ at strength $0.5$ \tabularnewline
\hline
\raggedright $(\mu,m_0)=(0.3,0.2)$ & \raggedright $1$ & \raggedright four levels per wedge & \raggedright between $3\times10^{-7}$ and $5\times10^{-7}$ at $R=30$ and $40$ & \raggedright linear \tabularnewline
\hline
\raggedright $(\mu,m_0)=(0.3,0.2)$ & \raggedright $2$ & \raggedright $(1,1)$ & \raggedright zero modes & \raggedright --- \tabularnewline
\hline
\raggedright $(\mu,m_0)=(0,0.2)$ & \raggedright $1$ & \raggedright four levels per wedge & \raggedright $4.6\times10^{-3}$ (extra sublattice symmetry at $\mu=0$) & \raggedright --- \tabularnewline
\hline\hline
\end{tabular}
\caption{Corner levels of the $(-,-)$ phase on wedges $|x|+|y|/\beta\le R$ ($R=40$ unless stated). The
first two rows use $t=1$, $t_y=0.15$, $a=0.8$, $b_1=0.3$, $b_2=0.13$, the last three $t=1$, $t_y=0.45$, $a=0.8$, $b_1=0.3$, $b_2=1$.}
\label{tab:corners}
\end{table}

\subsection{The double flip: intrinsic corners}
One crossing is pinned, so the parity is odd. At the double-flip point, every wedge ($\beta=1/2$, $1$, $2$,
with sharp or blunted tips) binds exactly one Majorana per mirror-axis corner, in the same sector at both
corners. The next level lies at $0.09$--$0.15$. The splitting of the two Majoranas decays exponentially
with $R$, from $8\times10^{-10}$ at $R=25$ to $10^{-14}$ at $R=40$ for $\beta=1$. Neither the symmetric corner
perturbation at strength $0.5$ nor mirror-symmetric disorder up to $w=1$ moves them. Breaking $W$
uniformly up to $\lambda=0.1$ does not split them either, because a single Majorana cannot gap locally
and leaves the corner only when an edge gap closes. The double flip is therefore an intrinsic second-order topological superconductor
protected by the reflection, of the kind induced by in-plane fields in helical
superconductors~\cite{Zhu2018,Pahomi2020}. It is also where the Euler class of \cref{app:model} is
odd.

\subsection{Fine-tuned corners on diamonds}
\begin{figure}[h]
    \centering
    \includegraphics[width=\textwidth]{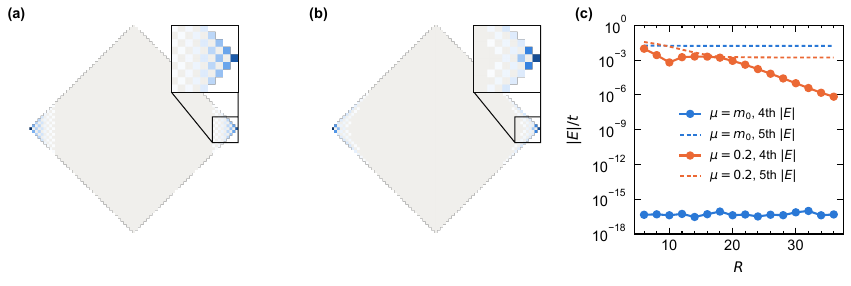}
    \caption{Diamond flakes $|x|+|y|\le R$ at $t=1.2$, $t_y=0.85$, $a=0.8$, $b_1=0.5$, $b_2=2$. (a),(b)~Summed
    density of the four lowest states at $R=36$, normalized to its maximum, (a)~at $\mu=m_0=0.05$ and
    (b)~at $\mu=0.2$, $m_0=0.05$; gray: the flake. Insets: zoom on the corner $(R,0)$. (c)~Fourth- and fifth-lowest $|E|$ versus $R$.}
    \label{fig:diamond}
\end{figure}

Corner modes can also appear without a topological reason. At the diamond point of \cref{tab:params}, which
lies in the $(-,-)$ phase, we cut diamonds $|x|+|y|\le R$. We searched over operators of the form (site
map)$\otimes$(sublattice sign)$\otimes\Gamma$, with $\Gamma$ any $4\times4$ matrix and the site map any of
the identity, $x\to-x$, $y\to-y$ and inversion, unitary or antiunitary, commuting or anticommuting. The
search finds only $\Xi$, $W$ and $\A$. Since $\mu=m_0$, the sublattice structure of \cref{app:invariants}
applies. For every $R$ from $6$ to $36$, the diamond has exactly four zero modes ($|E|<10^{-16}$), with
the next level at $0.017$. All four are in the same $W$ sector, with $63\%$ of their weight within five
sites of $(\pm R,0)$ and none near $(0,\pm R)$ [\cref{fig:diamond}(a,c)]. Every mode has
$\tau_z\sigma_y=+1$ and lives on one sublattice. In that block the on-site term $(m_0-\mu)\tau_z$
vanishes, and the remaining nearest-neighbor bonds anticommute with $(-1)^{x+y}$ and bind zero modes to
the sublattice excess at each corner. The $b_1$ and $b_2$ bonds that leave the block annihilate them
exactly. Off the line $\mu=m_0$, at $\mu=0.2$, the four lowest levels still approach zero
[\cref{fig:diamond}(b,c)]. Their splitting decays from $9\times10^{-4}$ at $R=20$ to $7\times10^{-7}$ at
$R=36$, with the next level at $0.0017$, and from $R=20$ on they share one sector and sit at $(\pm R,0)$.
Neither set is protected. A $\Xi$- and $W$-symmetric random perturbation within three sites of $(R,0)$
splits them linearly. With strength $\epsilon=10^{-2}$ at $R=28$, it gives $5\times10^{-5}$--$8\times10^{-4}$
at $\mu=m_0$ and raises the fourth level at $\mu=0.2$ from $2.6\times10^{-5}$ to
$6\times10^{-5}$--$3.5\times10^{-4}$, for four samples each. Removing the one to three outermost rows at
each tip leaves the exponential decay intact at $\mu=m_0$ ($10^{-11}$--$10^{-10}$ at $R=40$). At
$\mu=0.2$ it stops the decay, which saturates at $(7$--$9)\times10^{-5}$ for $R=32$ and $40$. Away from
$\mu=m_0$, the convergence is therefore a property of the single-site tip.

\section{Protection beyond the clean limit}
\label[appendix]{app:transport}

This appendix supports \cref{sec:protection}. We describe the Kwant calculation of the Andreev
conductance and apply it to the flat-band phase, with and without disorder. We then examine which edge
levels $\A$ protects on finite flakes, and how $W$ protects the four Majoranas of the $(-,-)$ phase.

\subsection{Andreev conductance}
A normal lead measures the number of Majoranas at an end through resonant Andreev
reflection~\cite{LawLeeNg2009, Flensberg2010, Wimmer2011}. We compute the zero-bias conductance $G=(e^2/h)(N_e-\sum|r_{ee}|^2+\sum|r_{he}|^2)$~\cite{BTK1982} from
the scattering matrix with Kwant~\cite{Groth2014}. We do so at the reference point and the reference $(-,-)$ point, with a
spot-check at the main-text couplings at the end of the next subsection. On flakes, a
semi-infinite normal lead (hoppings $t_L=4$ along $x$ and $1$ along $y$, $\mu_L=0$) covers the $x=0$ edge.
It couples to each edge site through a hopping $t_c=0.2$, which broadens the edge levels by
$\gamma\sim t_c^2/t_L\approx0.01$. This broadening is below the splitting by $\A$-breaking terms and the
dispersion of the $(-,-)$ edge band, so the conductance still distinguishes the flat band. For
$t_c\gtrsim0.4$ it no longer does. We evaluate $G$ at $E=10^{-5}$, on the plateau of $G(E)$. Exactly at
$E=0$, Majorana pairs split by about $10^{-7}$ produce a dip narrower than any physical resolution.

\subsection{Flat band: conductance and disorder}
On a flake of width $L_y$, the flat band becomes $2L_y$ zero modes on the $x$-normal edges. Contacting
one $k_y$ channel of a ribbon gives $G(0,k_y)=2.00\,e^2/h$ at every edge momentum [\cref{fig:transport}(b)].
On a flake with a wide lead, the zero-bias conductance is extensive~\cite{IkegayaSuzukiTanakaAsano2016},
$G(0)=2L_ye^2/h$ for $L_y=3$--$27$, with exactly $2L_y$ zero modes. Breaking $\A$ with $\lambda=0.2$
reduces it to $0.37\times2L_y$ at the largest widths [\cref{fig:transport}(c)]. On-site disorder uniform in
$[-w,w]$ leaves $G(0)/2L_y\ge0.998$ up to $w=1.25$, $2.5$ times the bulk gap, and $0.69$--$0.72$ at $w=2$,
whether or not it respects the mirror [\cref{fig:transport}(d)]. Without translation along the edge, $\A$
protects only the parity of the Majorana number~\cite{QueirozSchnyder2014, IkegayaKobayashiAsano2018},
which the localizer measures. With mirror-symmetric disorder on a $24\times9$ flake, the clean localizer
indices survive in every sample up to $w=1.5$ and in at most two of ten samples at $w\ge2$
(\cref{tab:locdis}), where the median localizer gap has fallen from $0.48$ to below $0.05$.

At the flat-band point $(\mu,m_0)=(2.4,2)$ (bulk gap $0.36$), $G(0,k_y)=2.00\,e^2/h$
at every $k_y$ tested. A $40\times L_y$ flake gives $G(0)/2L_y=0.90$, $0.94$ and $0.97$ for $L_y=5$, $9$
and $15$, reduced to $0.22$, $0.14$ and $0.11$ by $\lambda=0.2$. On $40\times9$ flakes (five samples),
disorder of strength $w=0.5$ and $0.7$ gives $0.89\pm0.01$ and $0.85\pm0.02$ when generic and
$0.90\pm0.02$ and $0.86\pm0.04$ when mirror-symmetric. The conductance per row is lower than at the
reference point because the edge modes decay more slowly into the bulk. The levels at the four corners
are then split by $3\times10^{-3}$--$5\times10^{-3}$, comparable to $\gamma$, and contribute only partly to $G(0)$.
This deficit becomes negligible as $L_y$ grows, and the conclusions are the same. A $(-,-)$ wire at
$(0.3,0.2)$ (wire gap $0.48$) carries two Majoranas per end, with $G(0)=4.00\,e^2/h$ and $\det r=+1$.

\subsection{\texorpdfstring{What $\A$ protects on flakes}{What A protects on flakes}}
\Cref{tab:edgelevels} gives the $2L_y$ lowest levels of rectangular flakes of $40$ columns at
$(\mu,m_0)=(2.4,2)$. The next level lies at $0.38$. Eight of the $2L_y$ levels do not converge, with $97\%$ of their weight within four sites of the four corners, where
the flat-band edges meet the gapped $y$-normal edges. The other $2L_y-8$ levels carry no weight there.
The corners break translation along the edge, and the levels next to them are not protected. A uniform
$0.05\,\tau_y\sigma_y$, which breaks $\A$ and keeps translation, raises the median to $1.5\times10^{-2}$
($L_y=41$). On-site disorder $V\tau_z+J'\tau_0\sigma_y$, with $V$ and $J'$ uniform in $[-w,w]$, gives a
median of $1.1\times10^{-3}$ at $w=0.2$ and $L_y=41$ (three samples) when it is symmetrized under
$y\to-y$, which keeps $W$ and $\A$. It gives $1.0\times10^{-3}$ when it is not symmetrized. On
$60\times21$ flakes, exactly two levels lie below $10^{-8}$ in every sample for $w=0.05$--$0.5$, with or
without the mirror. These are one Majorana per end, the parity of $\Xi$ for odd $L_y$. Without
translation along the edge, $\A$ therefore protects no more than this parity.

\begin{table}[h]
\centering
\begin{tabular}{lccc}
\hline\hline
$L_y$ & $11$ & $21$ & $41$ \\
\hline
median of the $2L_y$ lowest $|E|$ & $1.0\times10^{-4}$ & $1.2\times10^{-6}$ & $4.0\times10^{-8}$ \\
eight corner levels & --- & \multicolumn{2}{c}{$2.7\times10^{-3}$ and $4.6\times10^{-3}$} \\
\hline\hline
\end{tabular}
\caption{Lowest levels of clean $40\times L_y$ flakes in the flat-band phase, $(\mu,m_0)=(2.4,2)$.}
\label{tab:edgelevels}
\end{table}

With the tilt $\theta$ of the exchange defined in \cref{sec:protection}, a tilt of $\theta=5^\circ$
at $(2.4,2)$ widens
the edge band of a ribbon of $120$ columns to $0.020$, $0.025$ and $0.018\,\Delta_{\max}$ for pairing
amplitudes $(a,b_1,b_2)$ scaled by $1$, $0.3$ and $0.1$, with bulk gaps of $0.28$--$0.30\,\Delta_{\max}$.
The median disorder splitting above, $1.0\times10^{-3}$ at $w=0.2$, is $7.8\times10^{-4}\Delta_{\max}$. The
corresponding numbers for the adatom lattice are given in \cref{app:shiba}.

\subsection{\texorpdfstring{The $(-,-)$ phase}{The (-,-) phase}}
\begin{figure}[h]
    \centering
    \includegraphics[width=\textwidth]{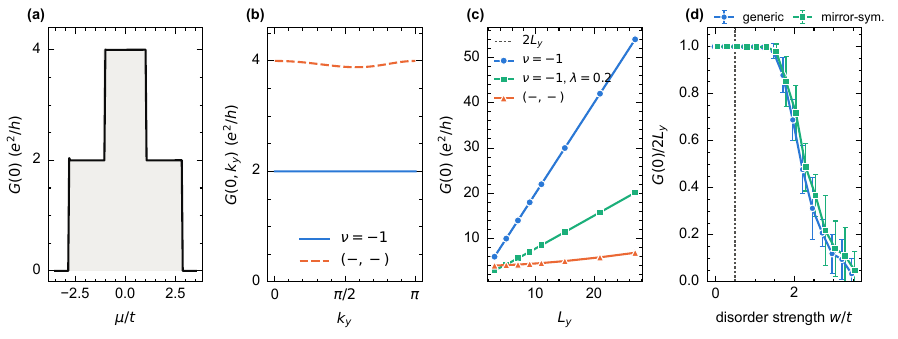}
    \caption{Andreev conductance at $t=1$, $t_y=0.15$, $a=0.8$, $b_1=0.3$, $b_2=0.13$. (a)~Wire ($L=120$, $m_0=0.9$):
    zero-bias conductance versus $\mu$, with the prediction $2\times$(number of inverted sectors)
    shaded. (b)~One $k_y$ channel of a ribbon
    ($N_x=60$): $G(0,k_y)$ at $(\mu,m_0)=(1.5,1.5)$ (flat band, $\nu=-1$) and $(-0.5,0.3)$ [$(-,-)$]. (c)~Flake ($L_x=40$) with a normal
    lead covering the $x=0$ edge (coupling $t_c=0.2$, $G$ at $E=10^{-5}$): $G(0)$ versus edge length $L_y$ at $(1.5,1.5)$,
    also with $\A$ broken by $\lambda\tau_y\sigma_y$, and at $(-0.5,0.3)$; dashed: $2L_y$.
    (d)~Disorder at $(1.5,1.5)$ ($40\times9$ flake, on-site disorder uniform in $[-w,w]$,
    $10$ samples, mean and standard deviation), generic and symmetric under $y\to-y$. Dashed: bulk
    gap $0.49$.}
    \label{fig:transport}
\end{figure}

\begin{figure}[h]
    \centering
    \includegraphics[width=0.55\textwidth]{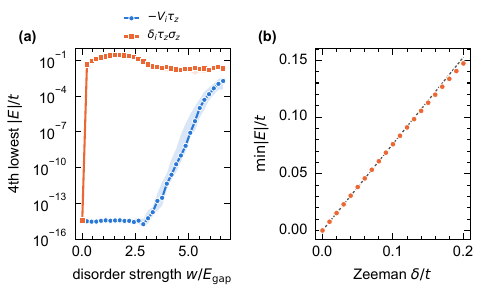}
    \caption{Protection of the four Majoranas of the $(-,-)$ phase by $W$ (strict wire, $L=120$, $t=1$, $t_y=0.15$, $a=0.8$, $b_1=0.3$, $b_2=0.13$,
    $(\mu,m_0)=(-0.5,0.3)$, wire gap $E_{\rm gap}=0.453$). (a)~Fourth-lowest $|E|$ versus disorder strength $w$ (uniform in
    $[-w,w]$, median and interquartile range over $20$ samples) for chemical-potential disorder $-V_i\tau_z$
    (blue), which commutes with $W=i\tau_z\sigma_y$, and for Zeeman disorder $\delta_i\tau_z\sigma_z$
    (orange), which anticommutes with it. (b)~Lowest $|E|$ versus a uniform Zeeman tilt
    $\delta\tau_z\sigma_z$; dashed: linear fit for $\delta\le0.07$, with slope $0.76$.}
    \label{fig:protection}
\end{figure}

On a wire of $L=120$ sites, the zero-bias conductance follows $0$, $2$ and $4\,e^2/h$ to within $0.05$ at
every gapped point of a sweep in $\mu$~\cite{DiezDahlhausWimmerBeenakker2012}. The scattering index
$\det r$~\cite{AkhmerovDahlhausHasslerWimmerBeenakker2011, FulgaHasslerAkhmerovBeenakker2011} is $-1$
only where the number of Majoranas per end is odd. It is therefore blind to
$(-,-)$, while the conductance is not [\cref{fig:transport}(a)]. On a flake with a wide lead, the conductance of the $(-,-)$ phase
starts near $4e^2/h$ and grows only slowly with $L_y$, to $6.9\,e^2/h$ at $L_y=27$, far below $2L_ye^2/h$
[\cref{fig:transport}(c)]. The modes are protected by $W$, which places one Majorana per end in each
sector [\cref{fig:protection}]. With $\Xi$ alone, the two at an end could pair up, since their total
class-D index is trivial. In a strictly one-dimensional wire, $W$ is the internal unitary
$i\tau_z\sigma_y$, which commutes with chemical-potential disorder. All four modes stay at zero to machine
precision in every sample up to a disorder strength of about $2.5$ times the wire gap. On-site Zeeman
disorder $\delta_i\tau_z\sigma_z$, which anticommutes with $W$ but respects $\Xi$, instead splits them at
any strength, and a uniform Zeeman tilt splits them linearly, $\min|E|\simeq0.76\,\delta$. The localizer
indices at the reference $(-,-)$ point survive mirror-symmetric disorder in a similar way (\cref{tab:locdis}).

\begin{table}[h]
\centering
\setlength{\tabcolsep}{10pt}
\begin{tabular}{lccccc}
\hline\hline
$w$ & $0.5$ & $1$ & $1.5$ & $2$ & $3$ \\
\hline
flat band, $(\mu,m_0)=(1.5,1.5)$ & $10$ & $10$ & $10$ & $1$ & $2$ \\
$(-,-)$, $(\mu,m_0)=(-0.5,0.3)$ & $10$ & $10$ & $9$ & $3$ & $2$ \\
\hline\hline
\end{tabular}
\caption{Number of $10$ samples in which the clean localizer indices survive mirror-symmetric on-site
disorder of strength $w$, on $24\times9$ flakes at $t=1$, $t_y=0.15$, $a=0.8$, $b_1=0.3$, $b_2=0.13$.}
\label{tab:locdis}
\end{table}

\section{Magnetic-adatom superlattice}
\label[appendix]{app:shiba}

This appendix details the magnetic-adatom model of \cref{sec:adatoms}. We map it onto \cref{eq:model},
show that its gap is odd under the mirror, describe the lattice and its Rashba coupling, and derive its
phase diagram, including denser chain geometries. We then read the phases on flakes with the
localizer, estimate the energy scales and tolerances, discuss spin-polarized tunneling, and list the
limitations of the model.

We call the model of this section a magnetic-adatom superlattice. The adatom moments are classical
spins, and the Shiba levels they bind are not projected onto \cref{eq:model}; the limitations of the model are discussed at the end of this appendix.

\subsection{Dictionary and mirror parity of the gap}
With $\epsilon_\bk=-\mu-2t\cos k_x-2t_y\cos k_y$,
$h(\bk)=\epsilon_\bk\sigma_0+J\sigma_y+\alpha(\sin k_x\,\sigma_y-\sin k_y\,\sigma_x)$ and
$\Delta(\bk)=D_x\sin k_x\,\sigma_x+iD_y\sin k_y\,\sigma_0$, the BdG Hamiltonian
$\left(\begin{smallmatrix}h(\bk)&\Delta(\bk)\\ \Delta^\dagger(\bk)&-h^*(-\bk)\end{smallmatrix}\right)$
in the basis $\tau\otimes\sigma$ equals \cref{eq:model} exactly (maximum deviation $0$ over $40$
random $\bk$), with
\begin{center}
\begin{tabular}{lll}
\hline\hline
term of \cref{eq:model} & microscopic origin & coefficient \\
\hline
$\epsilon_\bk\tau_z$ & normal dispersion & $t$, $t_y$, $\mu$ \\
$m_0\,\tau_0\sigma_y$ & exchange field along $y$ & $m_0=J$ \\
$b_1(\sin k_y\,\tau_0\sigma_x-\sin k_x\,\tau_z\sigma_y)$ & Rashba coupling & $b_1=-\alpha$ \\
$a\sin k_x\,\tau_x\sigma_x$ & triplet gap, $\mathbf d\parallel z$, $p_x$ & $a=D_x$ \\
$b_2\sin k_y\,\tau_y\sigma_0$ & triplet gap, $\mathbf d\parallel y$, $p_y$ & $b_2=-D_y$ \\
$a_1\sin k_x\,\tau_0\sigma_0$ (not used) & supercurrent, pair momentum $q$ & $a_1=2tq+O(q^2)$ \\
\hline\hline
\end{tabular}
\end{center}
The hole block must be $-h^*(-\bk)$. With Rashba coupling, $-h^*(\bk)$ differs from it and would break
particle--hole symmetry. Keeping only the $\sin k_y$ Rashba term would correspond to a Dresselhaus-type
coupling. The supercurrent term commutes with $H(\bk)$ and only tilts the spectrum. Bogoliubov Fermi
pockets~\cite{AgterbergBrydonTimm2017, YuanFu2018} open at $a_1=1.07$ at the reference point, about twice
the bulk gap, and at $a_1=0.43$ at the reference $(-,-)$ point.

Because $M_y^2=-1$, single-particle states carry mirror eigenvalue $\pm i$. A pair formed within one
sector is therefore mirror-odd, and a pair formed across the sectors mirror-even. For an even gap, the
BdG mirror commutes with $\Xi$, which then exchanges the two sectors. For an odd gap, it anticommutes with
$\Xi$, which then acts within each sector and makes each one a class-D
problem~\cite{UenoYamakageTanakaSato2013}. Consider pairing matrices
$A=(\mathbf d\cdot\boldsymbol\sigma)(i\sigma_y)$~\cite{SigristUeda1991} and the spin part $m=i\sigma_y$
of the mirror. Then $\mathbf d\parallel z$ and $\mathbf d\parallel x$ give $mAm^T=-A$ and need form
factors even in $k_y$, while $\mathbf d\parallel y$ gives $+A$ and needs one odd in $k_y$. The helical gap
of \cref{eq:model} is therefore odd. A singlet gap is even under every mirror, since
$m(i\sigma_y)m^T=i\sigma_y$ for all $m\in SU(2)$. In an $s$-wave Shiba lattice with moments along $y$, the
mirror that commutes with $H$ is $P_y\otimes\tau_0\otimes i\sigma_y$. It commutes with $\Xi$, so no
sector invariant exists. With out-of-plane moments, the $s$-wave lattice is instead a chiral topological
superconductor~\cite{RontynenOjanen2015, LiNeupert2016}.

\subsection{Lattice and Rashba coupling}
Magnetic impurities in a triplet superconductor bind Shiba states whose spin
structure depends on $\mathbf d$~\cite{BalatskyVekhterZhu2006, Kaladzhyan2016}. The substrate is \cref{eq:model} on a square lattice with $t=t_y=t_s=1$, $m_0=0$, $a=D_x=0.5$,
$b_2=-D_y=-0.4$ and $b_1=-\alpha=-0.2$. Classical spins with exchange $J\tau_0\sigma_y$ sit on one site of
each $a_x\times a_y=2\times4$ supercell. On a ribbon open along $x$ and periodic along $y$ with period
$a_y$, the mirror is $M(k_y)=\tau_z\otimes P_y(k_y)\otimes i\sigma_y$. Here $P_y(k_y)$ maps $j\to-j$ and
attaches $e^{-ik_ya_y}$ to sites that wrap into the neighboring cell. The mirror satisfies
$MH(k_y)M^{-1}=H(-k_y)$, and $\A=\tau_xM^*$ satisfies $\A H(k_y)^*\A^{-1}=-H(k_y)$, both to
$5\times10^{-16}$ at generic $k_y$. Each sector of \cref{eq:model} holds one negative-energy band, while a
classical spin contributes one Shiba level. A literal reduction to \cref{eq:model} would therefore need
two Shiba levels per cell forming a mirror pair. We do not attempt it, and we classify the lattice
directly by its own sector indices.

Without Rashba coupling the lattice has a chiral
symmetry. The operator $\tau_x\sigma_z$ anticommutes with $\tau_z$, $\tau_0\sigma_y$, $\tau_x\sigma_x$,
$\tau_y\sigma_0$ and $\tau_0\sigma_x$, hence with the substrate, the exchange and the triplet gaps. At
$\alpha=0$ the lattice is therefore in class BDI, and its flat band could be the chiral one. Only the
Rashba term $\tau_z\sigma_y\sin k_x$ commutes with $\tau_x\sigma_z$. Over $20$ random $\bk$ of the
supercell, $\max\|\{\tau_x\sigma_z,H(\bk)\}\|$ is $0$ at $\alpha=0$ and $0.40$ at $\alpha=0.2$. All
adatom results use $\alpha=0.2$, where the mirror and $\A$ remain exact (deviation $2\times10^{-16}$).

\subsection{Phase diagram}
For \cref{fig:shiba}(b) we scan $\mu\in[-4.5,4.5]$ and $J\in[0,4]$ on a
$91\times41$ grid. At each point, a ribbon of $100$ substrate columns at $k_y=0$ and $\pi/a_y$ is split into the
eigenspaces $M=\pm i$. The lowest $n$ levels of a sector count as zero modes when they lie below
$10^{-2}$ and a factor $100$ below the next level. Half their number is the number of Majoranas per edge
in that sector, and its parity gives $\nu_s$ on that line. The phases then follow as in \cref{tab:Sphases}.
Points without such a separation ($13\%$ of the grid, almost all with bulk gap below $0.03$) are marked
unresolved. A fixed threshold is not adequate. With $60$ columns and $|E|<10^{-6}$, edge pairs that decay
slowly along $x$, split by $10^{-6}$--$10^{-5}$ at bulk gaps $0.1$--$0.3$, are missed, and gapped points
appear as single flips. With the separation criterion, no single flip has a bulk gap above $0.034$, as
the gap closing forced by a change of $\nu$ requires. The bulk gap is the minimum over a $41\times41$
supercell grid, and local minimization changes it by less than $10^{-3}$ where we refined it.

The bare substrate ($J=0$) is time-reversal symmetric and is a helical class-DIII superconductor. A weak
exchange turns it into the double flip, which covers most of the region $J\lesssim2$, $|\mu|<3.9$. The
double flip is gapped at $89\%$ of its grid points with $J>0$, with a largest gap of $0.20$. Near the
substrate band edges, at $|\mu|\gtrsim3.5$ and $J\approx2.5$--$4$, the adatoms drive the trivial substrate
into the flat-band phase. This phase is gapped at $93\%$ of its points, with a largest gap of $0.18$. The
sector swap is nodal almost everywhere, and only $22$ of its points have a gap above $0.03$, at most
$0.04$. At $\alpha=0$ it was gapped at $80$ points, with gaps up to $0.07$. The Rashba coupling thus nearly
closes this gapped swap, and we do not discuss it further.
The table below lists representative points, with gaps refined by local minimization and zero-mode
counts identical for $100$ and $160$ columns.
\begin{center}
\begin{tabular}{lcccc}
\hline\hline
$(\mu,J)$ & zero modes $(+i,-i)$ at $k_y=0$ & at $k_y=\pi/a_y$ & bulk gap & phase \\
\hline
$(-1.6,0)$ & $(2,2)$ & $(0,0)$ & $0.20$ & helical (time-reversal symmetric) \\
$(-1.6,0.6)$ & $(2,2)$ & $(0,0)$ & $0.18$ & double flip \\
$(-3,1)$ & $(2,2)$ & $(0,0)$ & $0.13$ & double flip \\
$(-4.25,3.25)$ & $(0,2)$ & $(0,2)$ & $0.17$ & flat band \\
$(-3,3.5)$ & $(2,0)$ & $(0,2)$ & nodal & sector swap \\
$(-4.25,2)$ & $(0,2)$ & $(0,0)$ & nodal & single flip \\
\hline\hline
\end{tabular}
\end{center}
Along $\mu=-4.25$, below the substrate band edge, the adatoms drive the transition into the flat-band
phase, as the next table shows ($160$ columns).
\begin{center}
\begin{tabular}{lccccccc}
\hline\hline
$J$ & $0$ & $1$ & $2$ & $2.5$ & $3$ & $3.5$ & $4$ \\
\hline
zero modes at $k_y=0$, $\pi/a_y$ & $0,0$ & $0,0$ & $2,0$ & $2,2$ & $2,2$ & $2,2$ & $0,0$ \\
bulk gap & $0.242$ & $0.084$ & $0$ & $0.070$ & $0.159$ & $0.145$ & $0.105$ \\
$\max_{k_y}\min|E|$ & --- & --- & --- & $8\times10^{-15}$ & $1\times10^{-14}$ & $8\times10^{-15}$ & --- \\
\hline\hline
\end{tabular}
\end{center}
At $J=2$ one sector is inverted on one mirror line only (single flip), and the bulk is nodal, as
$\nu(0)\ne\nu(\pi)$ requires.

\subsection{Dense chains}
The flat-band phase of the $2\times4$ lattice is a narrow window in $\mu$, which also makes it fragile to
potential disorder (below). In the four-band model it requires $m_0>2t_y$, so chains that are dense along
$x$ and weakly coupled along $y$ should need less exchange. We repeated the classification for the
supercells $1\times3$, $1\times4$, $1\times5$, $1\times6$, $2\times3$, $2\times4$ and $2\times6$, on a
$46\times20$ grid with $\mu\in[-4.5,4.5]$, $J\in[0.2,4]$ and ribbons of $100$ columns. For dense chains
($a_x=1$), the gapped flat band covers $112$--$148$ grid points, with bulk gaps up to $0.31$--$0.34$. For
$a_x=2$ it covers $38$--$56$ points, with gaps up to $0.17$--$0.18$. The double flip reaches gaps of
$0.19$--$0.20$ in all seven supercells. Since the window extends below the substrate band edge, we
resolved it on a finer grid in $\mu$ for three supercells.
\begin{center}
\begin{tabular}{ccccccc}
\hline\hline
 & \multicolumn{2}{c}{$2\times4$} & \multicolumn{2}{c}{$1\times4$} & \multicolumn{2}{c}{$1\times3$} \\
$J$ & width in $\mu$ & max gap & width in $\mu$ & max gap & width in $\mu$ & max gap \\
\hline
$1.5$ & --- & --- & $0.5$ & $0.15$ & $0.4$ & $0.14$ \\
$2$ & $0.2$ & $0.07$ & $0.9$ & $0.23$ & $0.9$ & $0.23$ \\
$2.5$ & $0.5$ & $0.16$ & $1.4$ & $0.28$ & $1.4$ & $0.29$ \\
$3$ & $0.8$ & $0.18$ & $1.9$ & $0.31$ & $1.9$ & $0.30$ \\
$3.5$ & $1.0$ & $0.16$ & $2.3$ & $0.32$ & $2.4$ & $0.30$ \\
$4$ & $0.9$ & $0.14$ & $2.4$ & $0.32$ & $2.7$ & $0.30$ \\
\hline\hline
\end{tabular}
\end{center}
Dense chains thus widen the window by a factor of two to three, nearly double the gap and lower the
onset from $J\approx2$ to $J\approx1.5$. At the best $1\times4$ point, $(\mu,J)=(-4.1,3.6)$ with bulk gap
$0.33$, a ribbon of $120$ columns is flat to $3\times10^{-13}$. The window lies at and below the substrate
band edge, $\mu\approx-4t_s$, where the substrate is nearly empty and a fixed triplet gap is least
justified. The same holds for the $2\times4$ lattice. The figures and the localizer and disorder tests of
this section use the $2\times4$ lattice.

\subsection{Flakes and the invariant on flakes}
The flakes of \cref{fig:shiba}(c,d) are built from the same substrate and adatom
blocks. At the adatom double-flip point $(\mu,J)=(-1.6,0.6)$, the diamond $|x|+|y|\le32$ has two zero modes
($|E|=6\times10^{-8}$), with $96\%$ of their weight within six sites of the corners $(\pm32,0)$. The next level is at $0.031$. The helical substrate ($J=0$) instead has sixteen states below
$0.02$, the lowest at $2.3\times10^{-3}$, along the whole boundary of the same diamond. Of their weight,
$61\%$, $83\%$ and $96\%$ lie within $2$, $4$ and $8$ sites of the edge. The flat-band $60\times33$
rectangle at $(-4.25,3.25)$ has $18$ levels below $1.4\times10^{-3}$ on its $x$-normal edges, followed by
a gap to $0.15$.

\begin{figure}[h]
    \centering
    \includegraphics[width=0.85\textwidth]{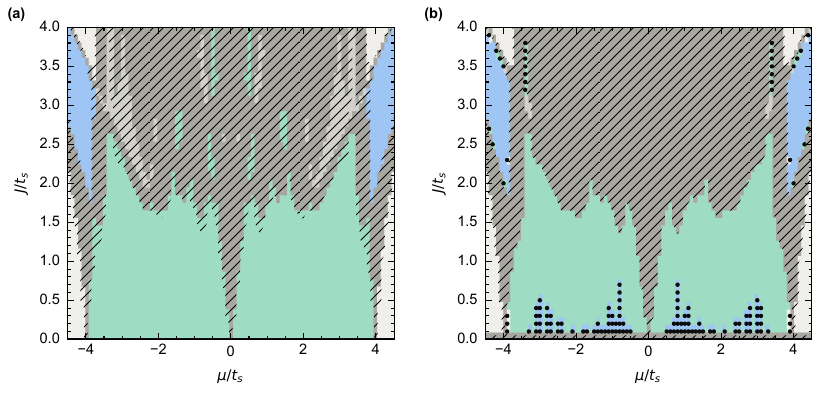}
    \caption{Phases of the adatom superlattice from two geometries. (a)~Ribbon classification of
    \cref{fig:shiba}(b): mirror-resolved zero modes of a ribbon of $100$ columns, open along $x$ and
    periodic along $y$; hatched: bulk gap below $0.03\,t_s$ or zero modes not resolved. (b)~Localizer
    classification: sector localizer at the center of two open $24\times17$ flakes, centered on and
    between adatom rows, $\kappa=0.1\,t_s$ per lattice spacing, read with \cref{eq:flakes}, on the same $91\times41$ grid, shown only
    where the ribbon is gapped and resolved (hatched otherwise). Thick line: the gapless helical line
    $J=0$. Dots: the $136$ of $1636$ points where the two readings differ, all with localizer gaps below
    $0.03\,t_s$, mostly at $J\le0.7$ next to the helical line.}
    \label{fig:shibaloc}
\end{figure}

The sector localizer applies to the open flakes directly. On an $L_x\times L_y$ rectangle centered on
the row $y=c$, the mirror about that row is a symmetry. As in \cref{eq:strip},
$\zeta=\zeta_+\zeta_-=\nu^{N_a}\,\nu_+^0\nu_+^\pi$, where the weak index now contributes a factor $\nu$
per adatom row ($N_a$ rows). The individual signs depend on the termination, but the product does not. We
use $L_x=24$, $L_y=17$ and $\kappa=0.1$, with one flake centered on an adatom row ($c=0$, $N_a=5$) and one
centered between two ($c=2$, $N_a=4$). They give $(\zeta_{\rm odd},\zeta_{\rm even})=(-,-)$ at the
adatom double-flip point $(-1.6,0.6)$, $(-,+)$ at the adatom flat-band point $(-4.25,3.25)$ and $(+,+)$ at
the adatom trivial point $(-4.4,0.3)$, with
sector-localizer gaps $0.05$--$0.45$. The four combinations $(\zeta_{\rm odd},\zeta_{\rm even})=(+,+)$,
$(-,-)$, $(-,+)$ and $(+,-)$ identify trivial or $(-,-)$, double flip, flat band and sector swap, relative
to the mirror about an adatom row. The mirror between two rows differs from it by a lattice translation
and exchanges the last two. On the $91\times41$ grid of the ribbon ($\mu\in[-4.5,4.5]$, $J\in[0,4]$, steps
$0.1$), this reading agrees with the ribbon classification at $1500$ of the $1636$ bulk-gapped, resolved
points with $J>0$ (\cref{fig:shibaloc}). All $136$ misses have localizer gaps below $0.03$ (median
$0.004$). Of these, $104$ lie at $J\le0.7$, next to the gapless helical line $J=0$, where the edge gap of
the flakes is small, and the remaining $32$ lie along phase boundaries at $J\ge2$. Where the localizer gap
exceeds $0.02$, the two readings agree at $1146$ of $1148$ points. This reading does not assume translation invariance.

We then added mirror-symmetric disorder, consisting of an on-site potential uniform in $[-w,w]$, a
relative spread of the adatom exchange uniform in $[-w/2,w/2]$, and missing adatoms with probability
$0.1$. All are drawn for $y\ge c$ and copied to $2c-y$. At the adatom double-flip point, the reading is the double
flip in all five samples at each of $w=0.25$, $0.5$ and $1$, the last about five times the bulk gap. At
the adatom flat-band point it is the flat band in $4$ of $5$ samples at $w=0.25$, $1$ of $5$ at $w=0.5$ and none
at $w=1$, mostly with localizer gaps below $0.04$. The flat-band phase of this lattice occupies a strip
narrower than $1$ in $\mu$ next to the substrate band edge. Potential disorder of the order of its gap
($0.17$) therefore moves parts of the flake out of the phase. The double flip covers most of
$|\mu|<3.9$ and is not affected.

\subsection{Energy scales and tolerances}
The parameters above use triplet amplitudes comparable to the hopping. To reach realistic ratios, we
scale $(D_x,D_y,\alpha)\to\lambda(D_x,D_y,\alpha)$ and compute the bulk gap, minimized from a $41\times41$
supercell grid, in units of $\Delta_{\max}=\max_\bk|\Delta(\bk)|=0.64\lambda$. The results are listed
below.
\begin{center}
\begin{tabular}{lcccccc}
\hline\hline
$\lambda$ & $1$ & $0.5$ & $0.3$ & $0.2$ & $0.1$ & $0.05$ \\
\hline
flat band, $J=3.25$ & $0.260$ & $0.255$ & $0.254$ & $0.253$ & $0.253$ & $0.253$ \\
double flip, $J=0.6\lambda$ & $0.283$ & $0.284$ & $0.284$ & $0.284$ & $0.284$ & $0.284$ \\
double flip, $J=0.6$ & $0.283$ & $0.226$ & $0$ & $0$ & $0$ & $0$ \\
\hline\hline
\end{tabular}
\end{center}
The gap never closes along the first two paths, so the phases persist to arbitrarily small triplet
gaps, with bulk gaps of about a quarter of $\Delta_{\max}$. For a triplet gap of $0.3$~meV the bulk gaps are thus about
$80~\mu$eV. The double flip requires the exchange to scale with the gap. At fixed $J=0.6$, it closes once $J$
exceeds two to three times $\Delta_{\max}$. The Rashba coupling must also stay below the gap. At fixed
$\alpha=0.2$, the flat band closes at $\lambda\le0.3$ and the double flip at $\lambda\le0.5$.

Two perturbations break $\A$ in practice. A tilt of the moments out of the substrate plane by an angle
$\theta$, which makes the exchange $J(\cos\theta\,\tau_0\sigma_y+\sin\theta\,\tau_z\sigma_z)$, breaks
$M_y$. An on-site singlet admixture $\lambda_s\tau_y\sigma_y$, even under every mirror, breaks the odd
parity of the gap. Both keep $\Xi$. The table below gives the width $\max_{k_y}\min|E|$ of the flat edge
band (ribbon of $100$ columns) and the lowest level of the double-flip diamond ($|x|+|y|\le32$).
\begin{center}
\begin{tabular}{ccc|ccc}
\hline\hline
$\theta$ & flat-band width & double-flip $|E_1|$ & $\lambda_s$ & flat-band width & double-flip $|E_1|$ \\
\hline
$0$ & $2\times10^{-11}$ & $6\times10^{-8}$ & $0$ & $2\times10^{-11}$ & $6\times10^{-8}$ \\
$1^\circ$ & $2.9\times10^{-4}$ & $6\times10^{-8}$ & $0.005$ & $2.1\times10^{-4}$ & $6\times10^{-8}$ \\
$2^\circ$ & $5.7\times10^{-4}$ & $6\times10^{-8}$ & $0.01$ & $4.2\times10^{-4}$ & $6\times10^{-8}$ \\
$5^\circ$ & $1.4\times10^{-3}$ & $5\times10^{-8}$ & $0.02$ & $8.5\times10^{-4}$ & $6\times10^{-8}$ \\
$10^\circ$ & $2.8\times10^{-3}$ & $4\times10^{-8}$ & $0.05$ & $2.1\times10^{-3}$ & $1\times10^{-8}$ \\
$20^\circ$ & $5.5\times10^{-3}$ & $1.2\times10^{-6}$ & $0.1$ & $4.2\times10^{-3}$ & $4\times10^{-7}$ \\
$30^\circ$ & $8.0\times10^{-3}$ & $5.7\times10^{-6}$ & & & \\
\hline\hline
\end{tabular}
\end{center}
The flat band acquires a width linear in the perturbation, $0.016\,\theta$ ($\theta$ in radians) and
$0.042\,\lambda_s$, just like the splitting by $\lambda\tau_y\sigma_y$ in \cref{eq:model}. The corner modes
of the double flip are single Majoranas $64$ sites apart, which can only leave zero energy by hybridizing
through the gapped edges. They therefore stay pinned. The next level drops from $0.031$ to $0.011$ at
$\theta=30^\circ$ and is unchanged by $\lambda_s$. Without Rashba coupling, the tilt preserves the chiral
operation $\tau_x\sigma_z$, which anticommutes with $\tau_z\sigma_z$. The flat band then stays flat to
$1.5\times10^{-10}$ at $\theta=30^\circ$, a direct sign of the chiral protection discussed above.

\subsection{Spin-polarized tunneling}
A tip polarized along the unit vector $\mathbf e$ measures, at zero bias,
$\sum_j|\langle\mathbf e|u_j\rangle|^2+|\langle\mathbf e|v_j^*\rangle|^2$, summed over the zero modes $j$
with particle and hole components $u_j$, $v_j$. On the mirror axis, a mode of sector $s$ is an eigenstate of $\tau_z\sigma_y$, and its zero-bias signal
is fully polarized along $y$. We find $P_y=-1.00$ for both corner modes of the double-flip diamond, and
$-0.82$ summed over the whole mode. In the flat-band rectangle, however, both sectors give the same
polarization, $P_y=-0.96$, set by the adatom moments. Spin-polarized tunneling therefore does not resolve
the sectors and is not a specific signature.

\subsection{Limitations}
\label[subappendix]{app:limits}
The spins are classical, the gap is not computed self-consistently, there is
one adatom per supercell, and the Shiba levels are not projected onto \cref{eq:model}. Three physical caveats apply. First, strong Rashba coupling favors triplet components with $\mathbf d$
parallel to the spin--orbit vector and mixes in singlet pairing~\cite{GorkovRashba2001, Frigeri2004,
Smidman2017}. The pairing of \cref{eq:model} is not of that form, so the continuum model assumes a Rashba
splitting below the triplet gap ($|b_1|=0.3<a=0.8$), and the adatom lattice uses $\alpha=0.2<D_x=0.5$.
The energy-scale analysis above shows that this condition is needed. Second, an exchange field along $y$
is pair-breaking for the $\mathbf d\parallel y$ component~\cite{SigristUeda1991}. In the adatom lattice
the exchange is local, but its effect on $D_y$ is not computed. The flat band does not need this
component, since it survives at $D_y=0$ (\cref{app:invariants}), and the exchange is perpendicular to the
$\mathbf d\parallel z$ component it does need. The double flip needs both. Third, substrate-mediated
coupling tends to order adatom moments into spirals~\cite{BrauneckerSimon2013, Klinovaja2013,
VazifehFranz2013, Christensen2016}, which break $M_y$. The in-plane collinear configuration is imposed
here, not derived.

\end{appendix}

\end{document}